\documentclass[useAMS,usenatbib,graphicx]{mn2e}
\usepackage{amsmath}
\usepackage{color}

\def\la{\mathrel{\hbox{\rlap{\hbox{\lower4pt\hbox{$\sim$}}}\hbox{$<$}}}}
\def\gt{\mathrel{\hbox{$>$}}}

\def\ga{\mathrel{\hbox{\rlap{\hbox{\lower4pt\hbox{$\sim$}}}\hbox{$>$}}}}
\def\lt{\mathrel{\hbox{$<$}}}

\usepackage{journal_names}
\usepackage{color}

\usepackage{url}
\usepackage[utf8x]{inputenc}

\newcommand{\fmg}{\mbox{\ensuremath{\;\mathrm{.}\!\!^\textrm{m}}}}
\renewcommand\fdg{\mbox{$.\!\!^\circ$}}

\newcommand\farcss{\mbox{$.\!\!\!^{\prime\prime}$}}
\def\farcm{\mbox{.\kern -0.5ex\raisebox{.6ex}{\scriptsize$\prime$}}}

\def\farcss{
 \mbox{ 
  \kern  0.13ex. 
   \kern -0.95ex\raisebox{.6ex}{\scriptsize$\prime\prime$}
  \kern -0.1ex
 }
}

\usepackage[hyperfootnotes=false]{hyperref}
\usepackage{breakurl}

\usepackage{graphicx,times}
\newcommand{\galfit}{\textsc{Galfit}}

\newcommand{\iraf}{\textsc{Iraf}}
\newcommand{\stsdas}{\textsc{Stsdas}}
\newcommand{\analysis}{\textsc{Analysis}}
\newcommand{\isophote}{\textsc{Isophote}}
\newcommand{\ellipse}{\textsc{Ellipse}}

\newcommand{\daophot}{\textsc{Daophot}}
\newcommand{\psf}{\textsc{PSF}}

\newcommand{\hi}{\textsc{Hi}}

\usepackage{float}

\title[The near-infrared $K_s$-band FP of the Norma cluster]{The Norma cluster (ACO\,3627) -- III.\ 
The Distance and Peculiar Velocity via the Near-Infrared $\bmath{K_s}$-band Fundamental Plane.}
\author[T.~Mutabazi et al.]
{\parbox{\textwidth}{T.~Mutabazi$^{1}$\thanks{E-mail: tom@ast.uct.ac.za},~
S.~L.~Blyth$^{1}$\thanks{E-mail: sarblyth@ast.uct.ac.za},~ 
P.~A.~Woudt$^{1}$\thanks{E-mail: pwoudt@ast.uct.ac.za},~
J.~R.~Lucey$^{2}$\thanks{E-mail: john.lucey@durham.ac.uk},~
T.~H.~Jarrett$^1$,~
M.~Bilicki$^{1}$,~ \\
A.~C.~Schr{\"o}der$^{3}$, ~and
S.~A.~W.~Moore$^{2}$} \vspace{0.4cm}\\
\parbox{\textwidth}{$^{1}$Astrophysics, Cosmology and Gravity Centre (ACGC), Astronomy Department, 
University of Cape Town, Private Bag X3, Rondebosch, 7701, South Africa\\
$^{2}$Department of Physics, University of Durham, South Road, Durham, DH1 3LE, United Kingdom \\
$^{3}$South African Astronomical Observatory, P.O. Box 9 Observatory 7935, Cape Town, South Africa} }

\begin{document}

\pagerange{\pageref{firstpage}--\pageref{lastpage}} \pubyear{2013}

\maketitle

\label{firstpage}

\begin{abstract}
While Norma (ACO\,3627) is the richest cluster in the Great Attractor
(GA) region, its role in the local dynamics is poorly understood.
The Norma cluster has a mean redshift (z$_{\rm{CMB}}$) of 0.0165 and has been proposed
as the ``core'' of the GA.
We have used the $K_s$-band Fundamental Plane (FP) to measure Norma cluster's distance with
respect to the Coma cluster. We report FP photometry parameters (effective radii and surface brightnesses), 
derived from ESO NTT SOFI images, and velocity dispersions, from AAT 2dF spectroscopy, for 31 early-type 
galaxies in the cluster. 
For the Coma cluster we use 2MASS images and SDSS velocity dispersion
measurements for 121 early-type galaxies to generate the calibrating
FP dataset. For the combined Norma-Coma sample we measure FP
coefficients of $a$\,$=$\,$1.465$\,$\pm$\,$0.059$ and
$b$\,$=$\,$0.326$\,$\pm$\,$0.020$. We find an rms scatter, in $\log
\sigma$, of $\sim$\,0.08 dex which corresponds to a distance
uncertainty of $\sim$\,28\% per galaxy. The zero point offset between
Norma's and Coma's FPs is $0.154$\,$\pm$\,$0.014$ dex. Assuming that 
the Coma cluster is at rest with respect to the cosmic microwave background frame 
and z$_{\rm CMB}$(Coma)\,=\,0.0240, we derive a distance to the Norma cluster of
$5026$\,$\pm$\,$160$\,km\,s$^{-1}$, and the derived peculiar velocity is
$-72$\,$\pm$\,$170$\,km\,s$^{-1}$, i.e., consistent with zero. 
This is lower than previously reported positive peculiar velocities for 
clusters/groups/galaxies in the GA region and hence the Norma cluster 
may indeed represent the GA's ``core''.
\end{abstract}


\section{Introduction} \label{sec_intro}

By studying the sky distribution of the Sc galaxy sample of \citet{Rubin_76} and the southern group catalogue of \citet{Sandage_75}, \citet{Chincarini_79} found evidence for a supercluster in the Hydra-Centaurus region. This supercluster encompasses the rich clusters now catalogued as 
Abell\;S0636 (Antlia, 
$v_{\rm{hel}}$\,$\simeq$\,2800\,km\,s$^{-1}$),
Abell\;1060 (Hydra, 
$v_{\rm{hel}}$\,$\simeq$\,3800\,km\,s$^{-1}$),
Abell\;3526 (Centaurus, 
$v_{\rm{hel}}$\,$\simeq$\,3400\,km\,s$^{-1}$) and
Abell\;3574 (IC4329, 
$v_{\rm{hel}}$\,$\simeq$\,4800\,km\,s$^{-1}$). The Local Supercluster has a sizable motion ($\sim$\,300\,km\,s$^{-1}$) towards this supercluster \citep{Shaya_84, Tammann_85} and the galaxy motions in the Local Supercluster have a shear that is directed towards Hydra-Centaurus \citep*{Lilje_86}. From a redshift survey of the Hydra-Centaurus region, \citet{daCosta_86} concluded that this supercluster extends to $v_{\rm{hel}}$\,$\simeq$\,5500\,km\,s$^{-1}$.

The discovery of large positive streaming motions in the Hydra-Centaurus direction \citep{Lynden_88} led to the idea of a large, extended mass overdensity, i.e., the Great Attractor (GA), dominating the local dynamics. They found a surprisingly large positive peculiar velocity for the dominant Cen30 sub-component of the Centaurus cluster, i.e., +1100 $\pm$ 208\,km\,s$^{-1}$. While several studies confirmed the presence of these large positive peculiar velocities, e.g., \citet{Aaronson_89, Dressler_90}, a non-significant peculiar velocity for Cen30 of +200 $\pm$ 300\,km\,s$^{-1}$ was measured by \citet{Lucey_88} (c.f.\ Burstein, Faber \& Dressler 1990).

Subsequently the SPS redshift survey \citep[supergalactic plane redshift survey,][]{Dressler_88} and the redshifts of IRAS galaxies in this region \citep{Strauss_88} confirmed that there was a substantial concentration of galaxies in this region extending from 2000 to 5500\,km\,s$^{-1}$. While in the literature the GA term has been used differently by various studies (see \citealt{Lynden_89, Burstein_90, Rowan_90, Mould_00, Courtois_12, Courtois_13}), an inclusive definition for the GA is the mass contained in the volume spanning $\ell$\,=\,260 to 350$^\circ$, $b$\,=\,--35 to 45$^\circ$, $v_{\rm{hel}}$\,=\,2000 to 6000\,km\,s$^{-1}$. This broad definition encompasses the Hydra-Centaurus supercluster but extends across the galactic plane to include Pavo-I, Pavo-II and the Norma cluster.

After more than two decades of study the nature and full extent of the GA is still not well established. The original work by \citet{Lynden_88} estimated the GA to have a mass of
$\sim$\,$5.4\times10^{16}\rm M_\odot$, centred at ($\ell$,\,$b$)\,$=$\,(307$^\circ$,\,9$^\circ$) and $v_{\rm{hel}}$\,$=$\,$4350$\,$\pm$\,$350$\,km\,s$^{-1}$. From SBF (surface brightness fluctuation) distances for $\sim$\,300 early-type galaxies (ETGs) \citet{Tonry_00} place the GA at ($\ell$,\,$b$,\,$v_{\rm{hel}}$)\,$=$\,($289^\circ$,\,$19^\circ$,\,$3200\pm260$\;km\;s$^{-1}$) with a mass $\sim8\times10^{15}\rm M_\odot$, i.e., approximately six times less than the original derived mass. As the GA spans low galactic latitudes, where the extinction is severe, our understanding of this important local large structure is still incomplete.

Beyond the GA at ($l$, $b$, $v_{\rm{hel}}$)\,=\,(312$^{\circ}$, 31$^{\circ}$, 14400\,km\,s$^{-1}$) lies the extremely rich Shapley Supercluster. While first recognised by \citet{Shapley_30} as a very populous cloud of galaxies, this structure was highlighted by \citet{Oort_83} in his review of superclusters. With the publication of the southern extension to the Abell cluster catalogue \citep*{Abell_89} a large number of rich clusters in Shapley was noted \citep{Scaramella_89}. \citet{Raychaudhury_89} from an analysis of galaxies on the UKST survey plates independently found this remarkable structure.  

The role of Shapley in the local dynamics is unclear. While most studies concluded/advocated that Shapley has only a modest contribution ($\leq$100\,km\,s$^{-1}$) to Local Group motion with respect to the CMB, e.g., \citet{Ettori_97, Branchini_99, Hudson_04}, some
studies have concluded a much larger contribution ($\sim$\,300\,\,km\,s$^{-1}$), see, e.g., \citet{Marinoni_98, Kocevski_06}. The lack of clear evidence for any backside infall into the GA
\citep{Mathewson_92, Hudson_94} led to the idea that the Shapley supercluster dominates the motions on the farside of the GA \citep{Allen_90}. Extensive redshift surveys (e.g., \citealt{Radburn_06, Proust_06}) 
have revealed the complex interconnections of the structures in this region.

The Norma cluster (Abell\;3627) was identified by \citet{Kraan_96} as the richest cluster in the GA region lying close to the galactic plane at ($\ell$, $b$, $v_{\rm{hel}}$)\,$=$\,(325$^\circ$, -7$^\circ$, \,4871\,km\;s$^{-1}$) \citep{Woudt_08}. Norma's mass and richness is comparable to the Coma cluster \citep{Woudt_08}. How the observed motions in the GA region relate to Norma is of 
great interest and this can be investigated via the measurement of Norma's peculiar velocity. In general, if Norma is indeed located near the core of the ``classical'' GA overdensity, then this cluster is likely to possess a small peculiar velocity.  Whereas if, for example, the relatively 
nearby GA model of \citet{Tonry_00} is correct, then the Norma cluster might be expected to
have a negative peculiar velocity of the order of --500\,km\,s$^{-1}$. Alternatively, if a sizable component of the observed GA flow were due to a large-scale bulk flow caused by the distant Shapley supercluster then the Norma cluster would possess a sizable positive motion of 
the order of +500\,km\,s$^{-1}$. The large-scale peculiar velocity field derived from density field 
reconstructions predict that Norma's peculiar velocity is less than 100\,km\,s$^{-1}$ (\citealt[][PSCz]{Branchini_99} and \citealt[][2MRS]{Lavaux_10}). 

There has been keen interest in the Norma cluster, with~dedicated multi-wavelength studies, for example X-ray \citep{Boehringer_96}, \hi\ \citep{Vollmer_01}, deep optical surveys \citep{Woudt_01}, a deep redshift survey \citep[][hereafter Paper~I]{Woudt_08} and near-infrared \citep*
{Skelton_09}.
 
In this third paper we report Norma's distance (and hence peculiar velocity) 
derived from distance measurements of early-type cluster galaxies. 
We use both the Fundamental Plane \citep{Djorgovski_87, Dressler_87_fp} 
and the metric aperture Faber-Jackson relation \citep{Faber_76, Lucey_86}. 
We apply these techniques using NIR $K_s$-band photometry where the effect 
of foreground galactic extinction is only $\sim$\,0.07 mag.
We determine Norma's distance relative to the Coma cluster which
we assume is at rest with respect to the CMB frame; various 
measurements support this assumption, e.g., \citep{Colless_01, Bernardi_02, Hudson_04, Springob_07}. 

This paper is structured as follows. Sample selection, observations,
and spectroscopic data analysis are discussed in
$\S$\;\!\ref{sample_data}. The photometric analysis is discussed in
$\S$\;\!\ref{data_analysis}. In $\S$\;\!\ref{offset}, we discuss the
methods used to determine the zero-point offset, and thereafter turn
the zero-point offset into a distance and peculiar velocity
($\S$\!\;\ref{distances}). We finally discuss our results under
$\S$\;\!\ref{sec_discuss}. 

We have adopted, where not stated, standard cosmology with $H_0 = 70.5\; \rm{km} \; \rm s^{-1} \!\;
\rm{Mpc}^{-1}$, $\Omega_\text{m}=0.27$, and $\Omega_{\Lambda}=0.73$
\citep{Hinshaw_09}. For the Coma cluster we adopt a redshift (z$_{_{\rm{CMB}}}$)
of 0.0240 which results in an angular diameter distance 
of 99.2\,Mpc and a scale of $1''$\,=\,0.481\,kpc for Coma.

\section{Sample Selection, Observations and Data Reduction} \label{sample_data} 

\subsection{Sample selection (photometry)}

\begin{figure}
 \centering
   \begin{tabular}{c c} 
       \includegraphics[width=0.46\columnwidth]{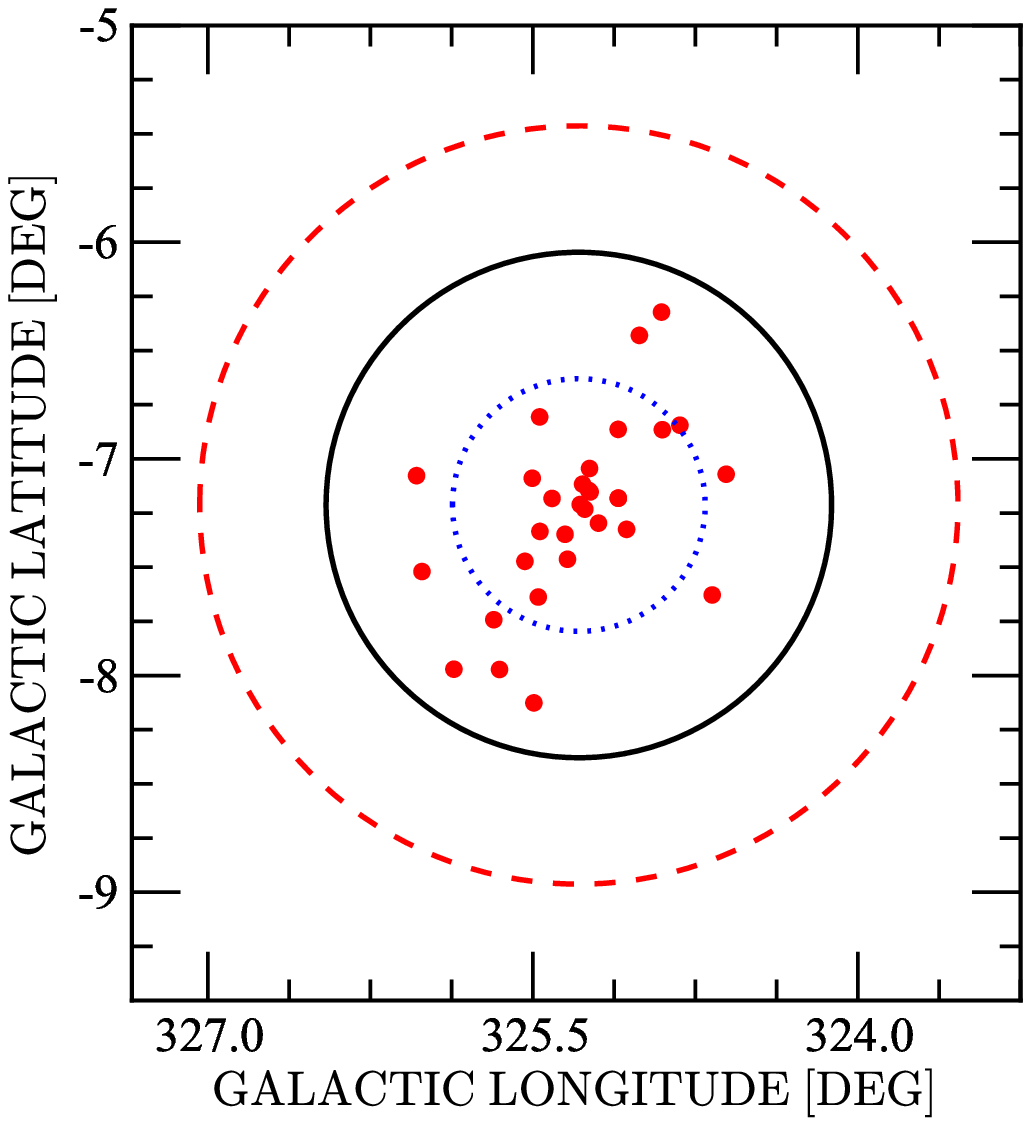}
       ~~
      \includegraphics[width=0.45\columnwidth]{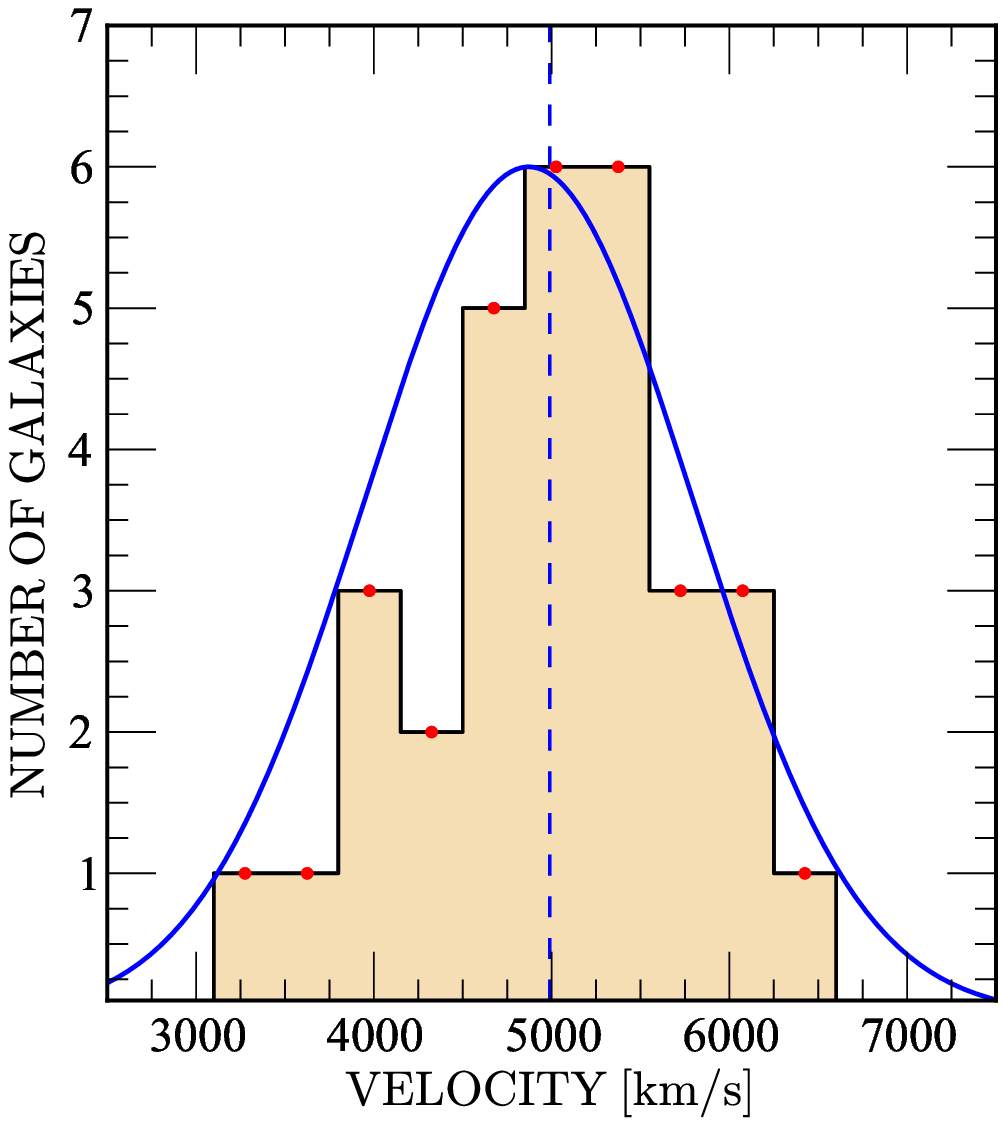} 
  \end{tabular}
\caption{
\label{sampleselec}
The sample selection for the Norma cluster NIR $K_s$-band FP. In the left panel, the outer dashed circle corresponds to the Abell radius, $R_A$, the black solid and the blue dotted circles correspond to $\frac{2}{3}R_A$ and $\frac{1}{3}R_A$, respectively. The red filled circles show the positions of the 31 ETGs. The right panel shows the corresponding velocity distribution; the over-plotted Gaussian (blue solid curve) corresponds to the mean heliocentric velocity and central velocity dispersion for the Norma cluster as measured from 296 cluster member galaxies (Paper I). The blue vertical dotted line represents the mean heliocentric velocity of our Norma cluster FP sample.}
\end{figure}

The Norma cluster has a mean heliocentric velocity ($v_{\rm hel}$) of
4871\,$\pm$\,54\,km\,s$^{-1}$ and a velocity dispersion ($\sigma_{\rm cl}$)
of 925\,km\,s$^{-1}$ (Paper I).
For our Norma ETG sample we selected galaxies that
\begin{enumerate}
\item had velocities within $\pm$\,3\,$\sigma_{\rm cl}$ of $v_{\rm hel}$,
i.e., the velocity range 2096\,km\,s$^{-1}$~$<v_{\rm hel}<$~7646\,km\,s$^{-1}$,
\item were within the Abell radius ($R_A$\,=\,1\fdg74), and
\item we had successfully measured a central velocity dispersion for
(see $\S$\ref{spec:data}).
\end{enumerate}
This resulted in a sample of 31 ETGs. Figure~\ref{sampleselec} (left panel)
shows the sky distribution of our sample; all galaxies lie within 0.6\,$R_A$.
Figure~\ref{sampleselec} (right panel) shows the velocity distribution of
the sample which is very similar to the distribution of the 296 known cluster members.

For our Coma cluster ETG sample we selected galaxies that
\begin{enumerate}
\item were typed as E or E/S0 or S0 by \citet{Dressler_80},
\item were redshift-confirmed cluster members,
\item had a central velocity dispersion in SDSS DR8
\citep{Aihara_11} and
\item we were successfully able to measure the FP photometry
parameters from the 2MASS Atlas images (see $\S$\,2.3).
\end{enumerate}
This resulted in a sample of 121 ETGs.

\subsection{Spectroscopic data} \label{spec:data}

Fibre spectroscopy was undertaken with the 2dF facility \citep{Lewis_02} on the 3.9\,m Anglo-Australian Telescope. For the Norma cluster three fibre configurations were used. This enabled repeat observations of Norma's ETGs for velocity dispersion measurements as well as an in-parallel redshift survey of the cluster (see Paper~I). Immediately before the observations of the Norma cluster, two fibre configurations centred on the Centaurus cluster (Abell 3526) were observed to calibrate the measured velocity dispersions onto a standardised system. Along with these science frames, observations were also made of four K-giant stars (to act as templates for the cross-correlation) as well as offset sky and flat-field frames.
 
The 2\,\farcs1 diameter of the 2dF fibres translates into a physical size of 0.70 kpc at the Norma redshift. As described in Paper~I, 
the observations were made with the 1200V gratings which resulted in a wavelength coverage of 4700--5840{\AA} at a FWHM resolution of 125 km\;s$^{-1}$ at Mg\,$b$; this is sufficient to determine velocity dispersions down to $\sim$\,60 km\;s$^{-1}$. Spectra were extracted from the raw data frames, flat-fielded, wavelength calibrated and sky-subtracted using the AAO {\tt 2dfdr} software package\footnote{\tt \url{http://www.aao.gov.au/2df/aaomega/aaomega_2dfdr.html}}. After redshift determination via cross-correlation, the spectra were shifted to a rest-frame wavelength and continuum subtracted. Velocity dispersions ($\sigma$) were measured by comparing galaxy spectra to those of the stellar templates, using the Fourier quotient method of \citet{Sargent_77}. Errors were determined by bootstrap re-sampling the spectra. 

Systematic offsets in velocity dispersion measurements at the level of $\sim$\,0.015\,dex exist between different observing systems (telescopes, spectrographs, runs, etc). The SMAC\footnote{Streaming Motions of Abell Clusters} project \citep{Hudson_01} intercompared measurements from 27 different systems and constructed a standardized system of velocity dispersions. Our Centaurus velocity dispersion measurements were used to calibrate our 2dF data onto the SMAC system. Table~\ref{cen_sigma_table} presents our velocity dispersion measurements for the Centaurus cluster galaxies; the independent measurements from the two different fibre configurations are reported. %
Prior to the comparison to SMAC, our measurements were averaged (weighted mean) and corrected to a standardized physical aperture size of $\sim$\,2\,$r_{\rm{norm}}$\;$=$\,1.19\,{\it h$^{-1}$} kpc following the prescription of \citet*{Jorgensen_95}: 
\begin{equation}
\log \sigma_{\rm{norm}} = \log{\sigma} + 0.04\,\log\left({\frac{r_{\rm{ap}}}{r_{\rm{norm}}}}\right); 
\end{equation}
where  $\sigma_{\rm{norm}}$ is the normalised velocity dispersion corrected to the standard aperture of radius $r_{\rm{norm}}$,  $\sigma$ is the measured velocity dispersion and $\sim$\,2\,$r_{\rm{ap}}$\,{\it
  h$^{-1}$} kpc for this work is the projected fibre diameter at the cluster distance. For our adopted cosmology (with $H_0 = 100h\; \rm{km} \!\; \rm s^{-1} \!\; \rm{Mpc}^{-1}$), at the Centaurus ($z=0.0123$), Norma ($z=0.0165$) and Coma ($z=0.0240$) distances, 2\,$r_{\rm{norm}}$ is equivalent to 6\,\farcs81, 5\,\farcs05, 3\,\farcs51, respectively; for fibre diameters of 2\,\farcs1, 2\,\farcs1 and 3$''$ the corrections are $-0.0204$, $-0.0152$ and $-0.0027$, respectively. 

\begin{figure}
\centering
\includegraphics[width=0.45\textwidth]{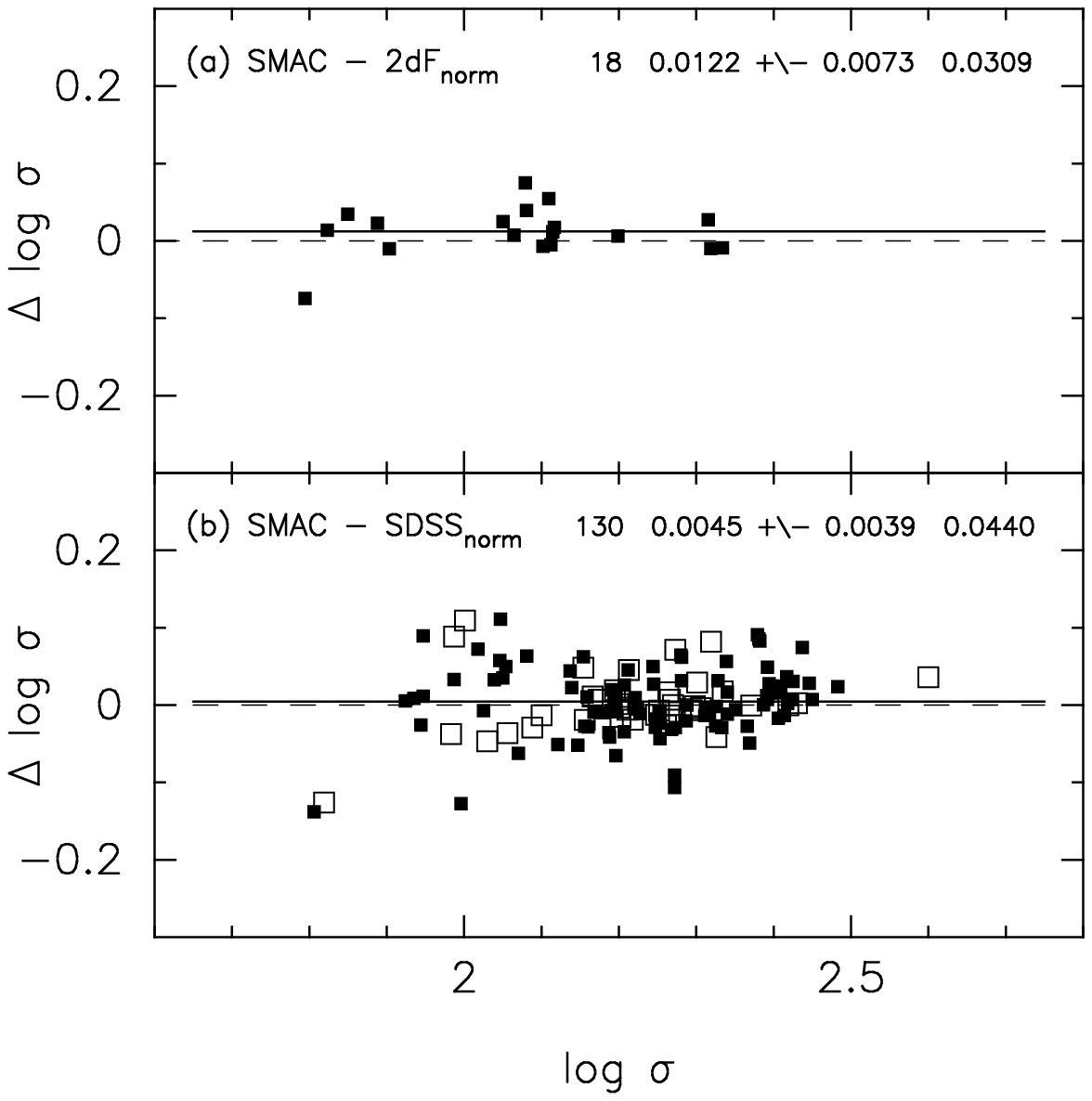} 
\caption{Comparison of the velocity dispersion measurements from SMAC, 2dF (aperture corrected) and SDSS (aperture corrected). The upper panel is the comparison of the 2dF Centaurus measurements with the SMAC values. The lower panel is the comparison of the SDSS and SMAC. Open squares represent the Coma cluster galaxies. 
}
\label{sigma_comparison}
\end{figure}

The two fibre configurations centred on the Centaurus cluster resulted in 26 velocity dispersion measurements for 18 galaxies. There is good agreement between our 2dF measurements and the SMAC standardized values with an observed mean offset (SMAC--2dF$_{\rm{norm}}$)  of +0.0122 $\pm$ 0.0073 dex  (see Fig.~\ref{sigma_comparison}, {\it upper panel}). There is a sizable overlap ($N=130$) between the SMAC survey and the more recent fibre velocity dispersion measurements in the northern hemisphere by the SDSS~DR8 \citep{Aihara_11
}. There is also good agreement between these two surveys with an observed mean offset (SMAC--SDSS$_{\rm{norm}}$) of +0.0045 $\pm$ 0.0039 dex (see Fig.~\ref{sigma_comparison}, {\it lower panel}). Included in this comparison are 38 Coma cluster galaxies; these have an offset of +0.0035 $\pm$ 0.0067 dex. The Coma cluster galaxies are represented by the open squares in the lower panel of Fig.~\ref{sigma_comparison}. 

Our velocity dispersion measurements for the Norma cluster galaxies are presented in Table~\ref{norma_sigma_table}. The individual measurements from the three fibre configurations are given.  A total of 112 velocity dispersion measurements were made of 66 galaxies of which we present only the 31 galaxies in our Norma sample. Also given in Table~\ref{norma_sigma_table} for each galaxy is the averaged (weighted mean) velocity dispersion corrected to the standardized physical aperture size and the SMAC system, i.e., a correction of {+0.012} dex is applied. The uncertainty in calibrating to the SMAC system is 0.007 dex which translates into systematic distance error at Norma of $\sim$\,2\%. 
\begin{table*}
 \centering
  \caption{Velocity dispersions for the Centaurus cluster. The $\sigma_1$ and
$\sigma_2$ columns refer to velocity dispersion measurements made with the two different fibre configuration.  $err_1$ and $err_2$ are the errors. $\sigma_{\rm{norm}}$ is the weighted average of the  $\sigma_1$ and $\sigma_2$ measurements with the aperture correction applied (--0.0204 dex) and $\log{\sigma_{_{\rm{SMAC}}}}$ is from table~7 of \citet{Hudson_01}.}
\begin{tabular}{lllrrrrccccl} 
\hline\hline
 Identification  & {RA (2000.0)} &  {DEC (2000.0)} &  $\sigma_1$ & err$_1$ &  $\sigma_2$ & err$_2$  & log $\sigma_{\rm norm}$ & err$_{\rm norm}$  & log $\sigma_{\rm SMAC}$ & err$_{\rm SMAC}$\\
\hline
E322-075 &  12 46 26.00 & --40 45 08.6  & 139.7 &  2.6 & 128.9 &  2.9  &  2.110  &   0.006 & 2.164 & 0.038 & \\
         &  12 48 31.02 & --41 18 24.1  &       &      &  83.9 &  5.0  &  1.903  &   0.026 & 1.893 & 0.026 & \\
         &  12 49 18.60 & --41 20 08.0  &  81.0 &  3.8 &       &       &  1.888  &   0.020 & 1.911 & 0.026 & \\
E322-099 &  12 49 26.27 & --41 29 22.6  & 121.6 &  2.1 &       &       &  2.065  &   0.008 & 2.072 & 0.026 & \\
E322-101 &  12 49 34.55 & --41 03 17.6  & 165.4 &  2.7 & 165.8 &  2.6  &  2.199  &   0.005 & 2.205 & 0.019 & \\
N4706    &  12 49 54.17 & --41 16 46.0  & 226.1 &  2.7 &       &       &  2.334  &   0.005 & 2.325 & 0.014 & \\
         &  12 50 11.54 & --41 13 15.8  & 113.8 &  3.2 & 119.5 &  2.2  &  2.050  &   0.007 & 2.075 & 0.009 & \\
         &  12 50 11.87 & --41 17 57.0  &  66.3 &  4.8 &  72.9 &  4.6  &  1.823  &   0.021 & 1.837 & 0.018 & \\
E323-005 &  12 50 12.26 & --41 30 53.8  & 219.0 &  3.2 & 215.2 &  2.7  &  2.316  &   0.004 & 2.343 & 0.026 & \\
E323-008 &  12 50 34.40 & --41 28 15.2  & 137.1 &  3.0 &       &       &  2.117  &   0.010 & 2.134 & 0.014 & \\
E323-009 &  12 50 42.98 & --41 25 49.5  & 136.4 &  1.9 &       &       &  2.114  &   0.006 & 2.126 & 0.018 & \\
         &  12 51 37.33 & --41 18 12.3  & 126.2 &  3.2 &       &       &  2.081  &   0.011 & 2.120 & 0.025 & \\
         &  12 51 47.97 & --40 59 37.4  &  74.0 &  3.5 &  74.2 &  3.1  &  1.849  &   0.014 & 1.884 & 0.025 & \\
         &  12 51 50.85 & --41 11 10.7  &  65.3 &  5.2 &       &       &  1.794  &   0.035 & 1.720 & 0.025 & \\
         &  12 51 56.51 & --41 32 20.2  & 132.9 &  3.4 & 132.3 &  2.7  &  2.102  &   0.007 & 2.095 & 0.025 & \\
N4743    &  12 52 16.02 & --41 23 25.8  & 135.7 &  2.3 &       &       &  2.112  &   0.007 & 2.107 & 0.021 & \\
         &  12 52 22.58 & --41 16 55.5  &       &      & 218.4 &  3.7  &  2.319  &   0.007 & 2.309 & 0.025 & \\
         &  12 52 40.86 & --41 13 47.3  & 124.1 &  5.1 & 127.3 &  4.9  &  2.079  &   0.012 & 2.154 & 0.025 & \\

\hline
\end{tabular}
\label{cen_sigma_table}
\end{table*}
%
\begin{table*}
\centering
\caption{Velocity dispersions for the Norma cluster. The $\sigma_1$, $\sigma_2$ and $\sigma_3$ columns refer to velocity dispersion measurements made with the three different fibre configuration, $\sigma_{c}$ is the weighted average of the $\sigma_1$, $\sigma_2$ and $\sigma_3$ measurements with the aperture correction ($
-0.0152$ dex) and the run offset of $+0.0122$ dex applied.} 
\begin{tabular}{lllrrrrrrccl} 
\hline\hline
 Identification  & {RA (2000.0)} &  {DEC (2000.0)} &  $\sigma_1$ & err$_{1}$ &  $\sigma_2$ & err$_2$ & $\sigma_3$ & err$_3$ &  log $\sigma_{\rm c}$& err$_{\rm c}$\\
\hline
WKK5920 & 16 07 52.618 & --60 31 12.95 &  206.8 &  4.6 &        &      &       &      &  2.312  &   0.010 \\
WKK5972 & 16 09 16.053 & --60 31 51.00 &  254.8 &  4.8 &        &      & 277.8 &  7.5 &  2.414  &   0.007 \\
WKK6012 & 16 10 12.103 & --61 16 01.15 &  151.3 &  5.2 &        &      & 147.1 &  6.4 &  2.172  &   0.012 \\
WKK6019 & 16 10 17.131 & --60 57 32.44 &  269.7 &  6.2 &  250.9 &  5.0 &       &      &  2.409  &   0.007 \\
WKK6047 & 16 10 58.873 & --60 55 24.51 &   96.8 &  3.7 &        &      & 120.9 &  6.3 &  2.010  &   0.013 \\
WKK6116 & 16 12 11.560 & --60 47 00.00 &  230.6 &  6.3 &        &      & 212.9 &  6.8 &  2.344  &   0.009 \\
WKK6180 & 16 13 32.141 & --61 00 22.69 &  212.4 &  6.5 &  200.6 &  4.9 &       &      &  2.308  &   0.008 \\
WKK6183 & 16 13 32.930 & --60 49 23.71 &  240.3 &  5.0 &        &      &       &      &  2.377  &   0.009 \\
WKK6198 & 16 13 54.073 & --61 37 57.00 &   81.0 &  5.1 &        &      &  81.1 &  5.1 &  1.905  &   0.019 \\
WKK6204 & 16 13 56.206 & --61 00 40.92 &  314.4 &  4.8 &        &      & 322.3 &  5.4 &  2.499  &   0.005 \\
WKK6221 & 16 14 02.807 & --60 29 41.07 &  116.0 &  5.8 &  105.3 &  5.2 &       &      &  2.038  &   0.015 \\
WKK6229 & 16 14 10.410 & --60 51 01.34 &  163.3 &  5.0 &        &      &       &      &  2.210  &   0.013 \\
WKK6233 & 16 14 18.007 & --60 53 25.87 &  176.1 &  5.0 &        &      &       &      &  2.243  &   0.012 \\
WKK6235 & 16 14 22.561 & --61 08 38.17 &        &      &        &      & 139.3 &  5.1 &  2.141  &   0.016 \\
WKK6242 & 16 14 30.538 & --60 53 46.35 &        &      &  280.5 &  6.8 & 249.6 &  7.8 &  2.424  &   0.008 \\
WKK6250 & 16 14 45.228 & --61 01 50.68 &  219.4 &  6.2 &  210.0 &  3.8 &       &      &  2.324  &   0.007 \\
WKK6269 & 16 15 03.833 & --60 54 25.61 &  381.9 &  6.9 &        &      &       &      &  2.579  &   0.008 \\
WKK6282 & 16 15 15.440 & --60 56 15.54 &  183.0 &  5.7 &        &      & 194.1 &  4.9 &  2.274  &   0.009 \\
WKK6305 & 16 15 32.922 & --60 39 55.27 &  220.3 &  4.0 &  207.9 &  3.8 &       &      &  2.327  &   0.006 \\
WKK6318 & 16 15 50.163 & --60 48 10.67 &  227.8 &  5.9 &        &      &       &      &  2.354  &   0.011 \\
WKK6342 & 16 16 18.929 & --60 57 23.60 &  234.9 &  6.0 &  206.7 &  3.4 &       &      &  2.326  &   0.006 \\
WKK6360 & 16 16 36.915 & --61 02 45.90 &  323.4 &  4.7 &  330.0 &  6.6 & 310.2 &  7.8 &  2.505  &   0.005 \\
WKK6383 & 16 17 00.358 & --60 52 24.96 &  165.3 &  4.9 &        &      & 147.9 &  5.7 &  2.195  &   0.010 \\
WKK6431 & 16 17 57.345 & --60 55 22.98 &  191.8 &  4.9 &        &      & 197.0 &  4.2 &  2.286  &   0.007 \\
WKK6473 & 16 18 41.300 & --60 17 38.00 &  112.6 &  4.6 &        &      & 130.5 &  8.3 &  2.064  &   0.015 \\
WKK6477 & 16 18 48.612 & --61 04 47.79 &  128.6 &  5.2 &        &      &       &      &  2.106  &   0.018 \\
WKK6555 & 16 20 15.031 & --61 00 44.75 &  161.6 &  3.3 &        &      & 162.9 &  3.9 &  2.207  &   0.007 \\
WKK6600 & 16 21 06.095 & --60 37 08.90 &  221.9 &  4.7 &        &      & 215.2 &  6.6 &  2.338  &   0.008 \\
WKK6620 & 16 21 26.857 & --61 11 06.12 &  57.1  &  9.9 &        &      & 64.2  & 10.9 &  1.777  &   0.053 \\
WKK6615 & 16 21 26.940 & --61 24 42.51 &        &      & 129.1  &  5.8 &       &      &  2.108  &   0.020 \\
WKK6679 & 16 22 34.872 & --61 02 14.19 &  141.4 &  3.4 &        &      & 155.0 &  5.5 &  2.159  &   0.009 \\

\hline
\end{tabular}
\label{norma_sigma_table}
\end{table*}

\subsection{Photometry: observations and data reduction}
 
While the effect of galactic extinction is significantly lower at NIR wavelengths than optical, the sky is also brighter. The NIR sky brightness also varies significantly at short time intervals. To avoid background saturation, NIR observations employ short exposures with a dithering pattern. Such short exposures ensure accurate sky determination, while the dithering mode minimises the effect of bad (dead, faulty) pixels. 

For the Coma cluster photometric analysis, we used the fully calibrated and reduced 2MASS Atlas images \citep{Jarrett_00}. 2MASS observations employed a total of six sky exposures each with 1.3\,s (total integration time $\sim$\,7.8\,s). The resulting frames (with 2$''$ pixel scale) were combined into Atlas images with resampled 1$''$ pixels. The average FWHM seeing was $\sim$\,3\,\farcs2. Detailed information about the 2MASS data reduction is given in \citet{Jarrett_00}.

The NIR imaging for the Norma cluster was conducted on four nights in June 2000 at the ESO, using the SOFI\footnote{\noindent SOFI (Son of ISAAC) is the infrared spectrograph and imaging camera on the ESO New Technology Telescope (NTT), covering 0.9\,--\,2.5~$\mu$m.} instrument on the 3.6\,m NTT. The SOFI imaging instrument provides a higher resolution due to the low pixel scale (0\,\farcs29 per pixel), and therefore provides higher quality (well resolved) images, than 2MASS. Such a relatively low plate scale imaging instrument, combined with good seeing conditions (mean FWHM for our observations is $\sim$~1\,\farcs08), is crucial, especially in the crowded, high stellar-density Norma region. For our sample, a total integration time of 300\,s split into $\sim$\,40 short exposures of 7.5\,s each was used. 

\begin{figure} 
 \centering
   \begin{tabular}{c c}
      \resizebox{38mm}{!}{\includegraphics[scale=1, bb = 30 273 581 660]{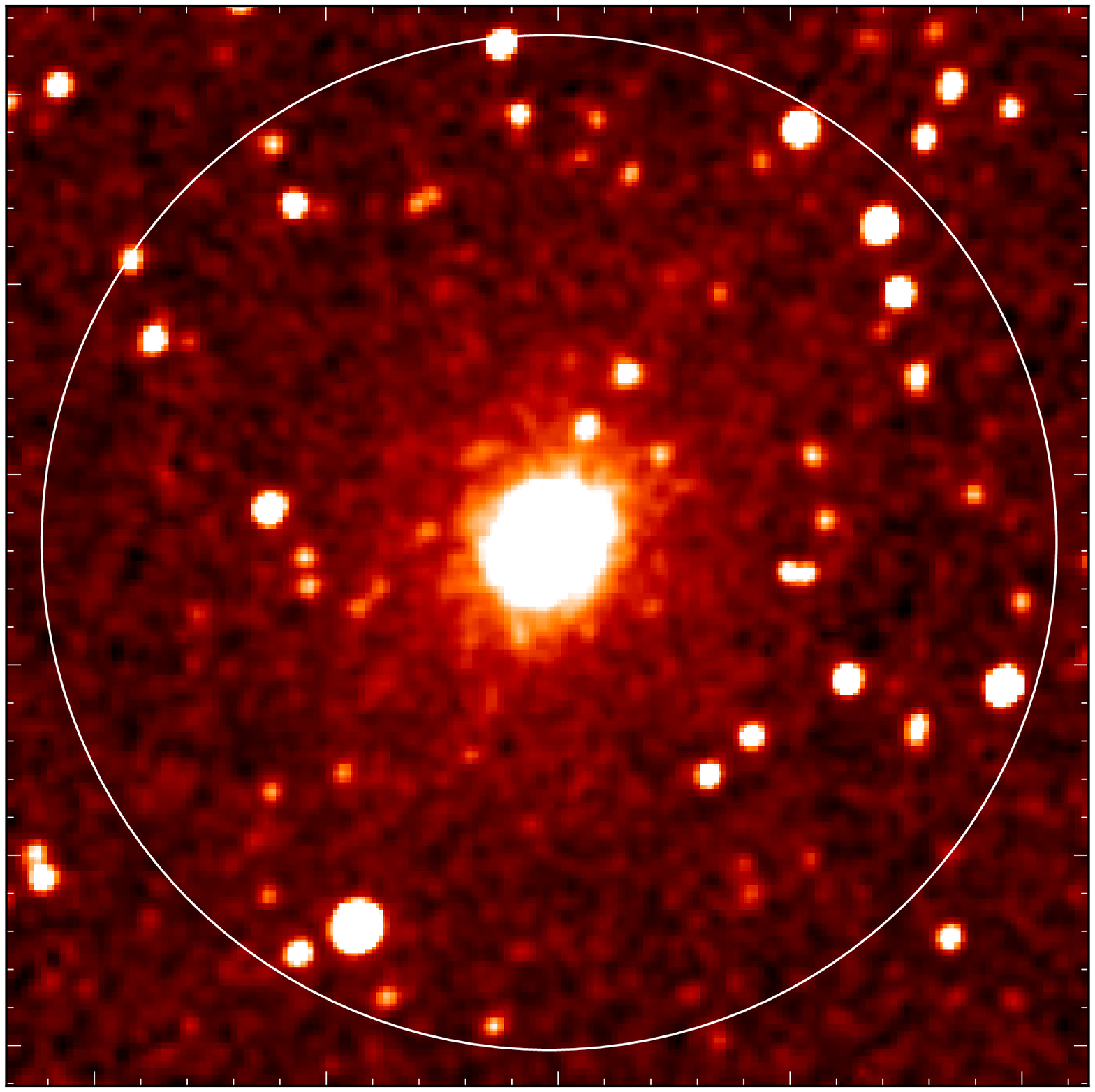}}
       \resizebox{38mm}{!}{\includegraphics[angle=270, scale=1, bb = 430 119 981 672]{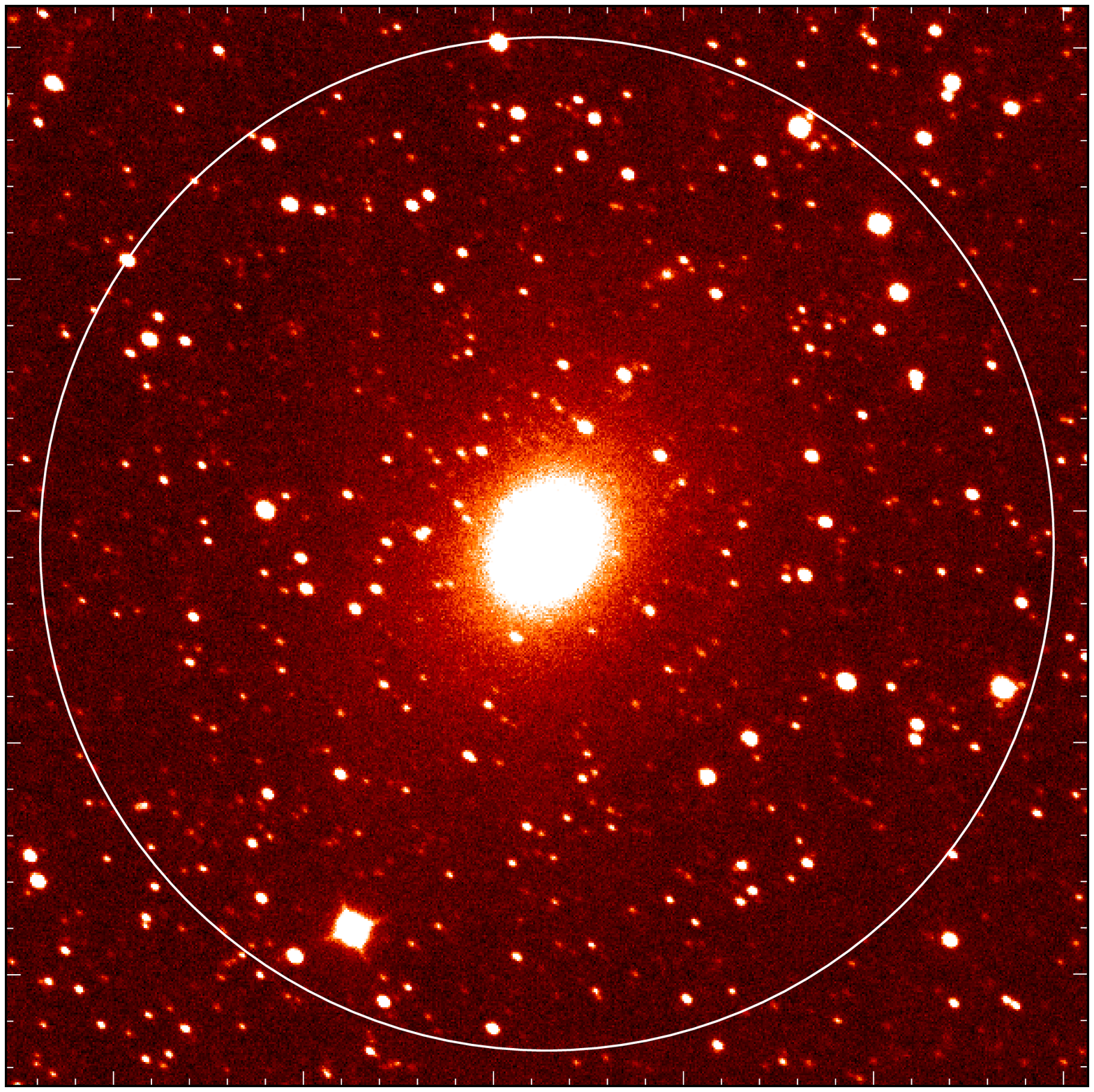}}
  \end{tabular}
 \vspace{-10em}
\caption[Comparison between 2MASS and ESO--NTT]{\label{2mass_compare}Images of one of the Norma cluster galaxies in our FP sample; WKK\,6318, from 2MASS (left panel) and ESO/NTT (right panel). 
}
\end{figure}

Standard NIR data reduction procedures were applied, including: dark subtraction, flat-fielding, sky-subtraction, and combining the dither frames for each target into a single science image. Image calibration (both astrometric and photometric zero-points) was performed using the 2MASS Point Source Catalogue \citep[2MASS~PSC, ][]{Skrutskie_06}. Figure~\ref{2mass_compare} shows a comparison between a 2MASS (left panel) and a SOFI (right panel) $K_s$-band image for the Norma cluster galaxy WKK\,6318. The white circle represents $80''$. Clearly, the low pixel scale of the SOFI imaging instrument coupled with our deep observations significantly improves the quality of the images and hence the reliability of the photometric results -- the high resolution of the SOFI instrument results in well resolved point sources which can then easily be subtracted.

\section{Photometry data analysis} \label{data_analysis}

The FP is the relationship between two photometric parameters (that is, the effective radius, $r_e$ and the mean surface brightness within that radius, $\langle\mu_e\rangle$ ), and the central velocity dispersion. In this section we describe the methods adopted to determine $r_e$ and
$\langle\mu_e\rangle$. The Norma cluster is located close to the galactic plane where stellar contamination is severe and therefore special techniques were needed to reliably subtract the foreground stars.

\subsection{Star-subtraction}

\begin{figure*}%
 \centering 
   \begin{tabular}{l c} 
 \includegraphics[width=0.48\textwidth]{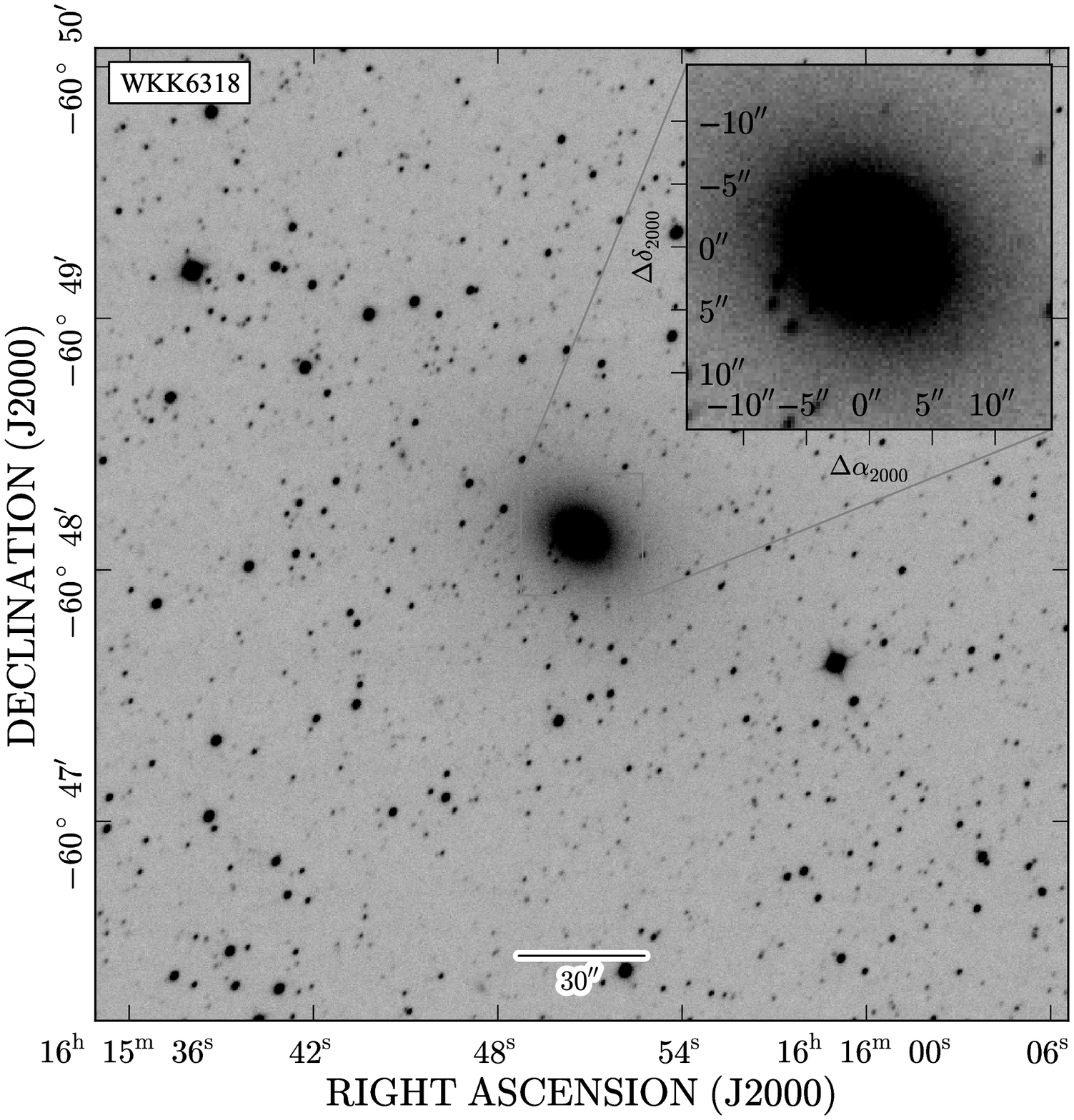} 
        \includegraphics[width=0.48\textwidth]{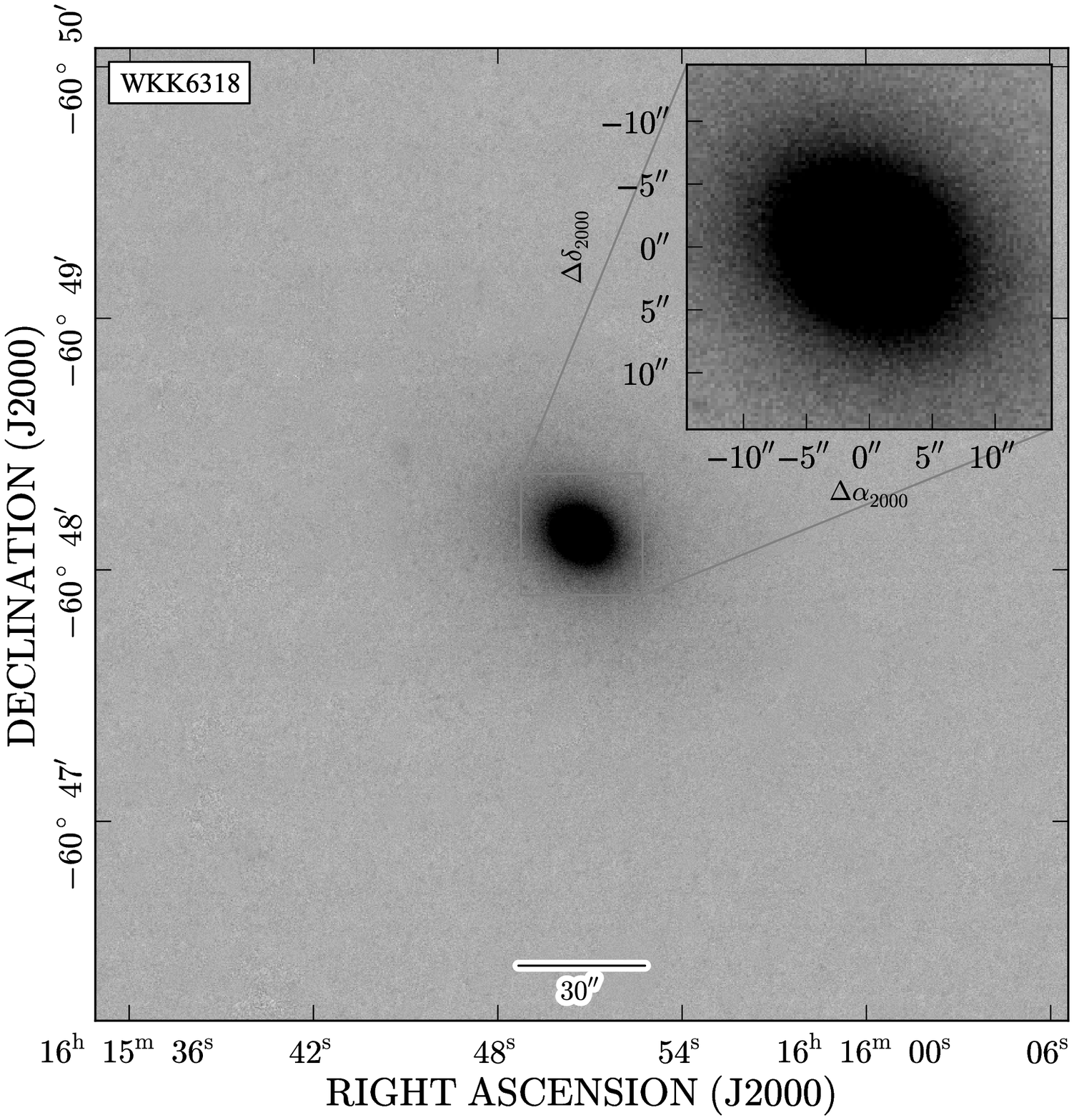} 
 \end{tabular} 
\caption{\label{kill6318} Star-subtraction -- WKK\,6318. The left panel shows the original image while the right panel shows the corresponding image after star-subtraction.}
\end {figure*}

Star-subtraction was performed using a script that employs various \iraf\footnote{\noindent \iraf\ is the Image Reduction and Analysis Facility; written and supported by the \iraf\ programming group at the National Optical Astronomy Observatories which are operated by the  Association of Universities for Research in Astronomy, Inc.\ under cooperative agreement with the National Science Foundation: \url{http://iraf.noao.edu/}} tasks mostly from the \iraf\ package \daophot\ \citep{Stetson_87}. Figure~\ref{kill6318} shows one of the images in our sample before and after star-subtraction (left and right panel, respectively). 
We have performed a thorough analysis to investigate the effect of our star-subtraction algorithm on our photometry for the Norma cluster. To quantify this effect, a simulation was performed using 12 ETGs from the Centaurus cluster. The choice of the Centaurus galaxies was motivated by the low stellar density. In the simulation, stars from a typical, randomly selected Norma field were added to the Centaurus galaxies. Photometric analysis was performed both before adding and after subtracting the superimposed stars. We found a small correction with a mean value of \mbox{$\bigtriangleup \text{m} = -0\!\fmg 0106 \pm 0\!\fmg 0003$} (see Fig.~\ref{av:mags}). We apply this value to correct the measured galaxy magnitudes for star-subtraction effects.

\subsection{Sky background estimation} \label{sky}

For accurate photometry, a reliable estimate of the sky background is crucial -- over-estimating the sky value results in galaxies appearing fainter than they are, and vice-versa.  We have determined the sky background within an annulus and measured the median sky value. The width of the annulus varied according to the initial estimate of the galaxy size, i.e., we approximated the galaxy size to be $\sim$~three times the measured effective radius ($r_e$) and set the inner and outer radii to $3.5\,r_e$ and $6\,r_e$, respectively. In cases (only for the Coma 2MASS images) where the outer radius ($6\,r_e$) is greater than the image size, the outer radius was set to match the image size but excluded the pixels on the edges. To minimise the effect of unresolved stars and other major artefacts, we iteratively applied a $2\sigma_s$ clipping where $\sigma_s$ is the standard deviation in the sky background within the annulus used. This effectively narrows down the effect of outliers at both the lower- and higher-ends of the pixel-value distribution. The resulting distribution is Gaussian in nature with the median sky value $\approx$ mean sky value.

\subsection{Determination of total magnitudes: surface brightness profile fitting} \label{ellipses}

Galaxy surface brightness profiles are usually fitted using a simple S\'{e}rsic \citep{Sersic_68} profile. In flux units, the single S\'{e}rsic component takes the form: 
\begin{equation}
I(r) = I_0 \exp \displaystyle{ \left[ - \left( \dfrac{r}{r_0} \right)^{\frac{1}{n}} \right] }. \label{sersic}
\end{equation}
$I_0$ refers to the central intensity, $r_0$ is the scaling radius, $n$ is the S\'{e}rsic index, also known as the concentration or shape parameter. Special surface brightness profiles, where $n$\,$=$\,$0.5$, $n$\,$=$\,$1$, $n$\,$=$\,$4$ are referred to as Gaussian, exponential and de~Vaucouleurs profiles, respectively. In units of magnitudes, Eq.~\ref{sersic} can be expressed in the form: 
\begin{equation}
 \mu(r) =  \mu(r_0) + 1.086 \displaystyle{ \left[ \left( \dfrac{r}{r_0} \right)^{\frac{1}{n}} \right] }; \label{sersic:used} 
\end{equation}
where $\mu(r_0)$ is the central surface brightness in mag\;arcsec$^{-2}$ and \mbox{$\mu_e \propto -2.5\, \log \, I_e$}. 

\begin{figure*}%
 \centering 
   \begin{tabular}{l c} 
   \includegraphics[width=0.48\textwidth]{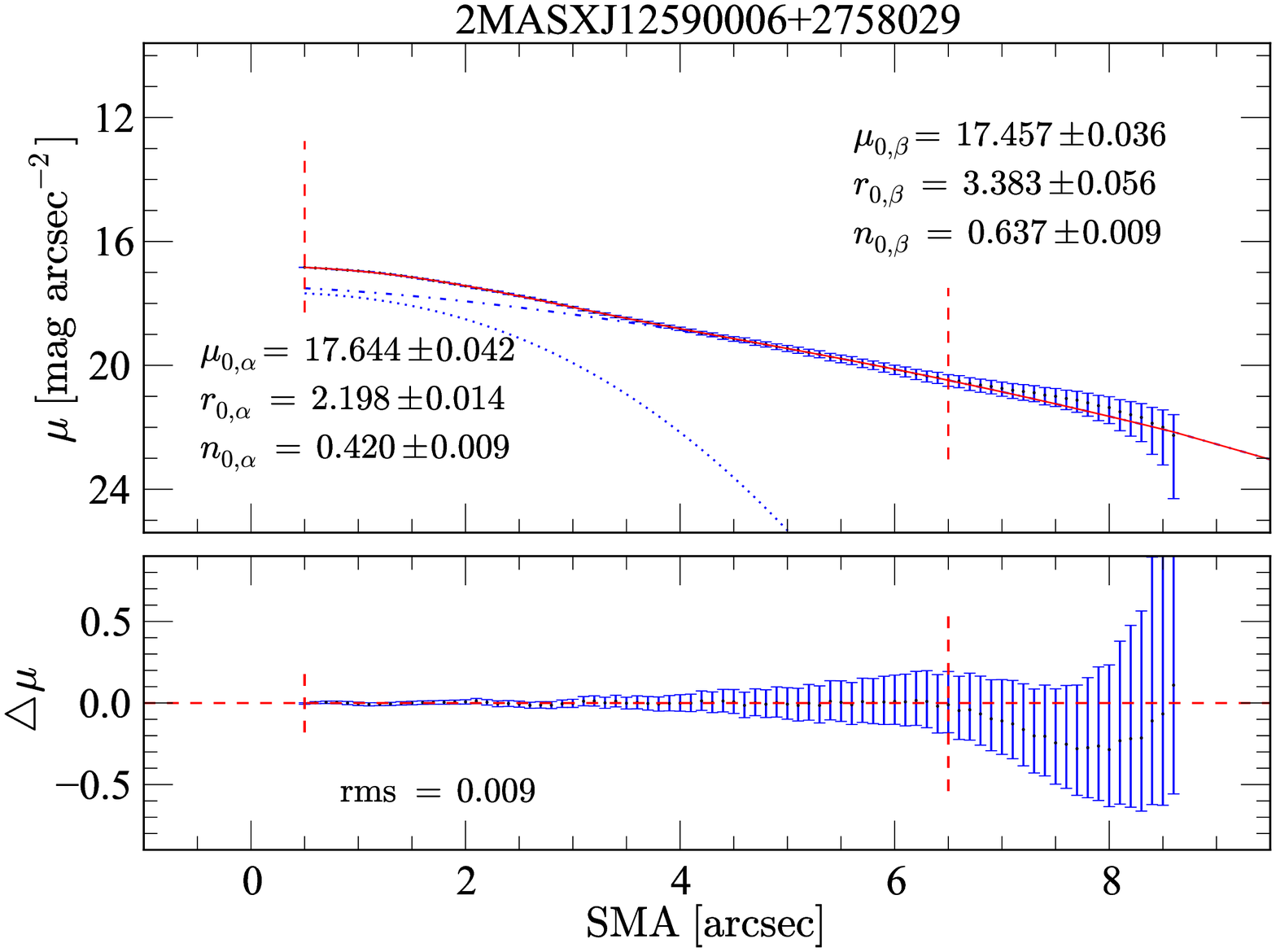}
\includegraphics[width=0.48\textwidth]{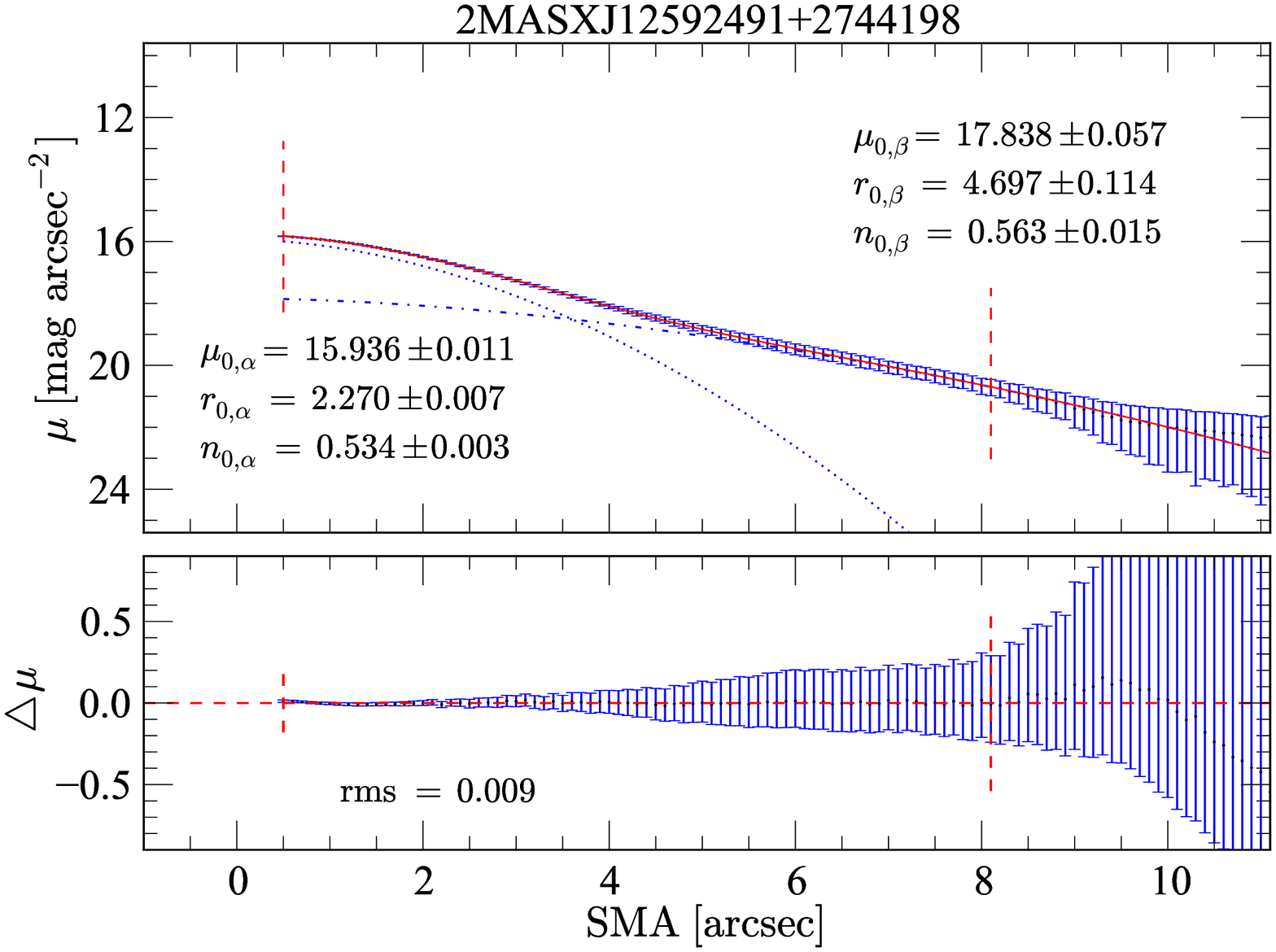} \\
 \includegraphics[width=0.48\textwidth]{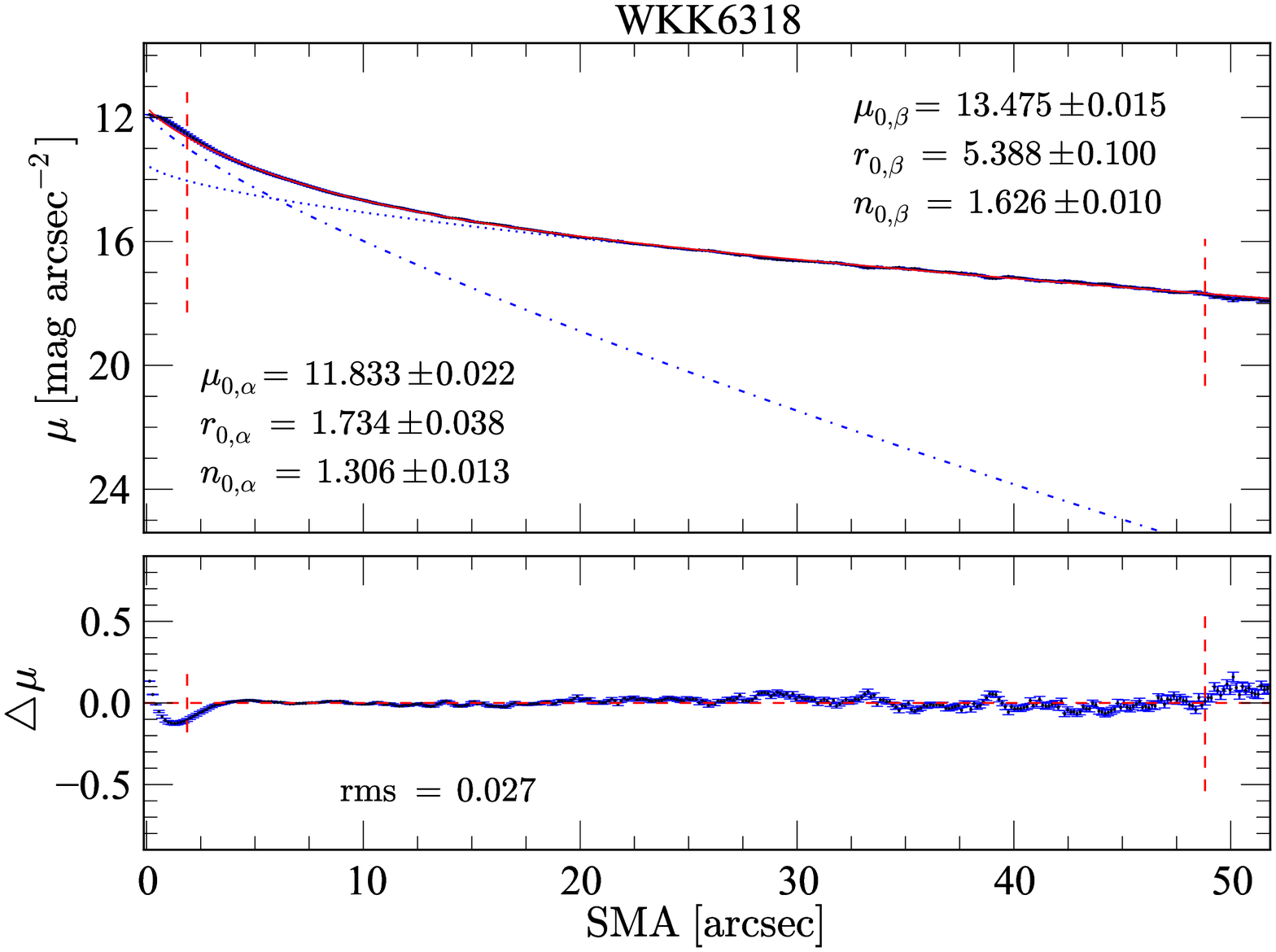}
 \includegraphics[width=0.48\textwidth]{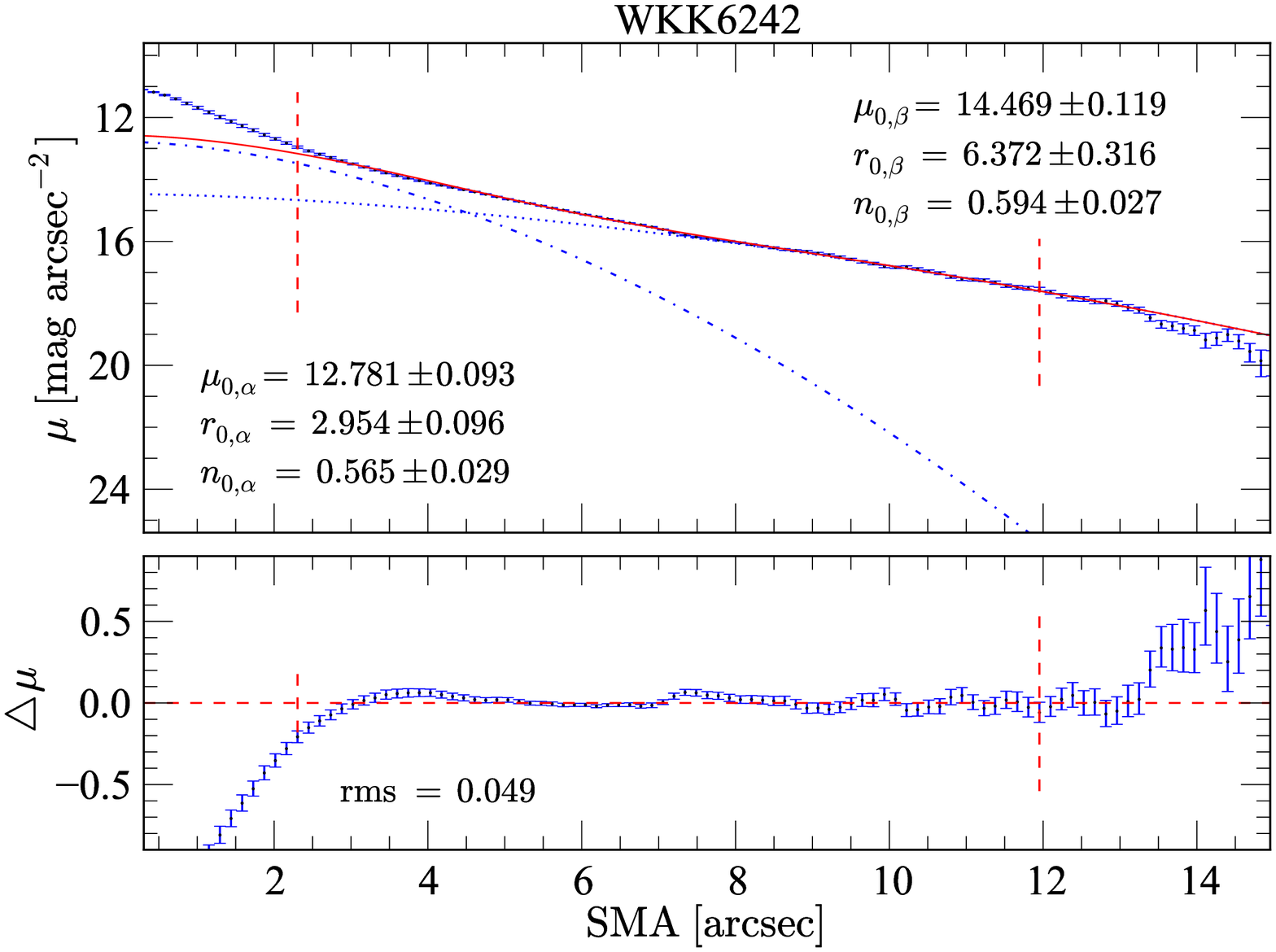}
 \end{tabular}
 %
\caption{\label{extrapolate:singDsersics_Nfit1d} Double S\'{e}rsic component fits to the galaxy surface brightness profile for some of the Coma (top left and right panels) and the Norma (bottom left and right panels) cluster galaxies in our sample. The dotted and dot-dashed blue lines represent the individual S\'{e}rsic components while the red solid line represents the combined fit. The small dashed vertical line on the extreme left corresponds to a radius twice the seeing (FWHM) for Norma and an inner radius of 0\,\farcs5 (for the Coma sample). The small dashed vertical line on the far right corresponds to the radius where the galaxy flux is approximately equal to the measured deviation in the sky background -- these two small vertical lines represent the range of the data points used to fit the galaxy surface brightness profile. The subscripts $\alpha$ and $\beta$ denote the fit parameters for the two separate S\'{e}rsic components which were used to simultaneously fit the galaxy surface brightness profiles. The best fit (solid red curve) is the combination of the two S\'{e}rsic components.}
\end{figure*}

In our photometric analysis, we have used the \ellipse\ task under the \stsdas--\analysis--\isophote\ package in \iraf, to fit isophotes and derive galaxy surface brightness profiles. The resulting surface brightness profiles were then fitted using a combination of two S\'{e}rsic functions (see, e.g., \citealt{Huang_13}) and the total flux was determined by extrapolation. The fitting method was applied to both Coma and Norma samples. Figure~\ref{extrapolate:singDsersics_Nfit1d} is an example of the fitted profiles, for Coma (top) and Norma (bottom). The red solid line represents the best fit (which is the sum of the two S\'{e}rsic components), which we extrapolate so as to determine the galaxy flux that would otherwise get lost within the background noise. The data used to fit the galaxy surface brightness profiles for the Norma sample were restricted for the radius ranging from twice the FWHM (indicated by the small dashed vertical lines on the extreme left) to where the galaxy flux is 1$\sigma_s$ above the sky background (vertical dashed lines on the extreme right). FWHM and $\sigma_s$ are the seeing and sky background deviation for each image, respectively. For the Coma sample, we restricted the data to within an inner radius of 0\,\farcs5 and an outer radius where the galaxy flux is 1$\sigma_s$ above the sky background.

The total luminosity, $L_{\rm tot}$ for a single S\'{e}rsic component, can be obtained by extrapolating the surface brightness profile to infinity. Within a given radius $r_{\rm max}$, the luminosity, $L_{r_{\rm max}}$ is given by integrating the galaxy profile to the given radius\footnote{For convenience, we have left out the factor $(1-e)$; where $e$ is the galaxy ellipticity, since it does not affect our final expression for the magnitude correction; $\rm{dm}$. The full expression should be: \\ $L_{\rm{tot}} = \displaystyle{ \int_0^{\infty} 2\pi (1-e) \, r I(r) \,\mathrm{d}r}$}. 
\begin{eqnarray}
L_{tot} &=& \displaystyle{ \int_0^{\infty} 2\pi r I(r) \,\mathrm{d}r} = 2\pi I_0 \,n\, {r_0}^2 \,\Gamma(2n) \label{sersic_single1};  \\ \nonumber \\ 
L_{r_{max}} &=& \displaystyle{ \int_0^{r_{max}} 2\pi r I(r) \,\mathrm{d}r}; \nonumber \\ 
&=&  2\pi I_0  \,n\, {r_0}^2 \, \displaystyle{\gamma\left[2n,\,\left(\dfrac{r_{max}}{r_0}\right)^{1/n} \right] }; \label{sersic_single2}
\end{eqnarray} 
where $\Gamma(x)$ and $\gamma(s, x)$ are the complete and incomplete gamma functions, respectively. %
For a double S\'{e}rsic profile, the total luminosity for the galaxy is computed by combining the two components. To recover the galaxy flux in the outer parts that would otherwise get lost within the background noise, we apply:
\begin{equation}
L_{\rm{tot}} = L\left(r = r_{\rm{max}}\right) + L\left( r_{\rm{max}}<r<r_{\infty} \right).
\end{equation}

\subsection{Effective radius and \psf\ correction} 

The effective (half-light) radius for each galaxy was measured from the circle enclosing half the total flux through interpolation. This was then corrected for seeing effects using \galfit\ \citep[Version 3.0.5;][]{Peng_10} following the description in \citet{Magoulas_12}, i.e., we used \galfit\ to model the galaxy images with and without \psf\ convolution. The effective radii were measured from the resulting \galfit\ models. The difference in the effective radius from the \psf\ convolved model ($r_{e,\rm{psf}}$) and the model without \psf\ convolution ($r_{e,\rm{nopsf}}$) is the seeing correction, i.e., $r_{e,\rm{psf}}-r_{e,\rm{nopsf}}$. This was subtracted from the effective radius measured from the original image. The effective surface brightness is computed from the seeing-corrected effective radius (circular) using \\
$\langle\mu_e\rangle = m_{_{\rm{tot}}} + 2.5\, \log (2\pi) + 5\, \log r_e$; \\
where $m_{_{\rm{tot}}}$ is the total extrapolated magnitude corrected for the star-subtraction and sky background effects. 

\subsection{Measurement errors and analysis}

The errors on the effective radius and the total extrapolated magnitude (and hence $\langle\mu_e\rangle)$ are correlated. For the Norma cluster sample, we modelled the errors in the photometric parameters using \galfit. Using the \galfit\ output parameters (total magnitude, effective radius along semi-major axis, axial ratio, position angle, S\'{e}rsic index) of the galaxy, we created mock images corresponding to the respective \galfit\ output parameters for each of the galaxies in our Norma cluster sample. Each of the mock galaxy models was convolved with the \psf\ image for the respective galaxy field before adding it to the different positions in the original galaxy image field (the data image) but avoiding the centre or the position of the original observed galaxy. Photometry was performed on each of the added fake galaxies and the output is compared with the input parameters. The difference between the simulated and true magnitude is negligible, with a median value of $-0\!\fmg008$. This indicates that our photometry is not significantly affected by sky gradients or faint unresolved stars in the field. By using the median value for each individual galaxy field, the measurement error on the total magnitude from the simulation is $\sim$\,0\!\fmg02\,$\pm$\,0\!\fmg01. In addition, we measured the magnitude error within an aperture whose size is four times the measured effective radius (taking into consideration the photometric zero-points error as well). The total error on the total extrapolated magnitude is the quadrature sum of the error obtained through the simulation 
and the aperture photometry error within the 4\!\;$r_e$ apertures. %
The measurement errors on the total magnitude for the 2MASS Coma cluster galaxies were determined following the description given in the 2MASS All-Sky Data Release Explanatory Supplement (see {\tt \url{http://www.ipac.caltech.edu/2mass/releases/allsky/doc/sec6_8a.html}}). 
The error on the total magnitude was calculated using:
\begin{equation} 
\bigtriangleup \rm{mag} = 1.087\, \dfrac{\left[\left(S/g_{\rm{av}}\right)+n\left(2k\sigma_s \right)^2+\left(0.024n\sigma_s \right)^2\right]^{1/2}}{S};
\end{equation} 
where $g_{\rm{av}}=G\times N$ where $G=10$ is the gain, $N=6$ is the total number of co-added frames, $S$ is the total flux, $k\approx 1.7$ is the Kernel smoothing factor, $\sigma_s$ is the co-added noise, and $n$ is the total number of pixels within an aperture. We used a size of aperture equal to four times the effective radius, so that the aperture magnitude within that radius would be a good approximation to the measured galaxy total extrapolated magnitude. 

For both Norma and Coma sample galaxies, we measured the corresponding error in the effective radius (arising from the error in the total magnitude), and applied simple propagation of errors to compute the error in the mean effective surface brightness. 
\begin{table} 
\centering
\caption{\label{m_errors}Corrections and systematic errors for the Norma ETGs. The extinction correction presented in this table is from SF11 (refer to section~\ref{g_extinc} for details). 
}
\begin{tabular}{lrcc} \hline\hline
     &   \multicolumn{1}{c}{Correction} & \multicolumn{1}{c}{Error} & S/M\\ 
     \multicolumn{1}{c}{(1)}            & \multicolumn{1}{c}{(2)}                & \multicolumn{1}{c}{(3)}         &  \multicolumn{1}{c}{(4)}           \\ \hline    
Star-subtraction [mag]                  &  \multicolumn{1}{c}{$-$0.0106}   &  0.0003   &   M \\
Sky background [mag]                 & \multicolumn{1}{c}{$-$0.008}    &  0.020   &    M   \\ 
Photometric calibration [mag]     & \multicolumn{1}{c}{-}                   &  0.025       &  S   \\ 
$\sigma$-aperture correction [dex] & \multicolumn{1}{c}{$-$0.015}  &  -  &   M   \\ 
$\sigma$-run offset [dex]               &  \multicolumn{1}{c}{0.012}        &  0.007  &  M   \\ 
Malmquist bias [\%]                         & \multicolumn{1}{c}{0.587}        &         -  &  S \\ 
Extinction correction [mag]             & \multicolumn{1}{c}{0.057 -- 0.088}   &        -  &   S   \\ 
$k$-correction [mag]                       &  \multicolumn{1}{c}{0.042 -- 0.071}     &         -  &   S \\ 
Seeing correction [$''$]                   &  \multicolumn{1}{c}{0.001 -- 2.607}        &         -  &  M \\ 
Cosmological dimming [mag]       & \multicolumn{1}{c}{ 0.051 -- 0.091}    &         -   &  S \\ 
\hline
\end{tabular}
\begin{flushleft}
\small{NOTES: \\ 
Column~1 represents the different corrections applied to the spectroscopic and photometric measurements for the Norma sample. Column~2 is the mean value of the correction. Where the correction was applied to individual galaxies, the range is given. Column~3 is the error on the mean value. M and S given in Column~4 refer to the type, i.e., measurement and systematic errors, respectively. 
}
\end{flushleft}
\end{table}

An independent photometric analysis was performed on the Coma sample, where the effective radius for each galaxy was measured based on the galaxy's total extrapolated magnitude from the 2MASS Extended Source Catalogue \citep{Jarrett_00, Skrutskie_06}. There is a small offset (median value) of $0\!\fmg 014\pm 0\! \fmg 009$ between these measurements and ours, with 2MASS magnitudes being brighter. 
The median difference in the X-component of the FP (i.e., $\log \,r_e - 0.32\langle \mu_e \rangle$), between our results and the independent measurements, is $-0.005\pm0.001$~dex. 

\subsection{Photometric corrections to $\bmath{\langle \mu_e \rangle}$ and systematics} \label{systematics}

\subsubsection{Galactic extinction corrections} \label{g_extinc}

The effect of galactic extinction can be corrected for using the DIRBE/IRAS reddening maps of \citet*{Schlegel_98}. It has been found however, that the \citeauthor{Schlegel_98} NIR reddening maps, over-estimate the extinction at low Galactic latitudes where they are uncalibrated -- see, e.g., \citet{Bonifacio_00, Schroder_07, Schlafly_11}. For our analysis, we adopt the correction factor of 0.86 by \citet{Schlafly_11} which is a modification of the \citeauthor{Schlegel_98} maps. 
We used NED\footnote{The NASA/IPAC Extragalactic Database (NED) is operated by the Jet Propulsion Laboratory, California Institute of Technology, under contract with the National Aeronautics and Space Administration.} to extract the \citet{Schlafly_11} extinctions in the Landolt $B$-band and converted them to the $K_s$-band using 
\mbox{$A_{K_s} = 0.085\,A_B$}. 
The extinction values for the Norma sample range from 0.057 to 0.088~mag (see Table~\ref{m_errors}). The mean difference due to the modification is, on average, 0\!\fmg011, which corresponds to a Norma distance offset of 45\,km\,s$^{-1}$.

\subsubsection{Redshift and cosmological dimming corrections to $\langle \mu_e \rangle$} 

We applied a $k$-correction, $k_{K_s}=-3.83z + 21.9z^2$, as given by \citet{Pahre_99}. Various redshift corrections exist, e.g., \citet{Bell_03} gives a $k$-correction of $-2.1z$. Applying this correction results in a magnitude difference of $0\!\fmg 02$ and $0\!\fmg 03$ for the Norma and Coma clusters, respectively. The corresponding distance offset for Norma is $\sim$\,33\;km\;s$^{-1}$. 

The cosmological dimming effect on the mean effective surface brightness, which is due to uniform expansion of space was corrected for, using the $(1+z)^{4}$ term -- the cosmological dimming correction term is $-10\log\,(1+z)$. At the redshift distances of Norma and Coma, the mean corrections are 0.07\;mag\;arcsec$^{-2}$ and 0.10\;mag\;arcsec$^{-2}$, respectively. 

\subsection{Fundamental Plane data: $\bmath{\log r_e}$, $\bmath{\langle \mu_e \rangle}$ and $\bmath{\log \sigma}$}
The FP relates the stellar properties (central velocity dispersion, $\sigma$), galaxy size (effective radius, $r_e$), and the mean galaxy surface brightness within the effective radius ($\langle\mu_e\rangle$) of ETGs, i.e., 
\begin{equation} 
\log r_e = a \log\sigma + b \langle \mu_e \rangle + c; \label{fp:2013}
\end{equation}
where $a$ and $b$ are the FP slopes, while $c$ is the intercept (zero-point). The fully corrected mean effective surface brightness ($\langle \mu_e \rangle$) that is finally used in fitting the FP is
\begin{equation}
\langle \mu_e \rangle = m_{\rm{tot}} + 2.5 \log \left( 2\pi r^2_e \right) - A_{K_s} + k_{K_s} - 10 \log \left( 1+z \right); 
\end{equation}
where $m_{\rm{tot}}$ is the measured total extrapolated apparent magnitude. 
Table~\ref{Norma_phot} shows the results for the Norma cluster sample. Included in the table are the final variables ($r_e$, $\langle \mu_e \rangle$, and $\log \sigma$) used to fit the FP. A similar table for the Coma cluster sample is provided in the appendix (refer to Table~\ref{Coma_phot}).
\begin{table*}
\begin{center}
\caption[Norma cluster photometry results]{\label{Norma_phot} Norma cluster photometry results obtained by f\mbox{}itting and extrapolating the galaxy surface brightness prof\mbox{}iles.}
\vskip-1.5em\begin{tabular}{c r r c c c c }  \\ \hline\hline
\multicolumn{1}{c}{Identification}  & \multicolumn{1}{c}{Tot. mag} &  \multicolumn{1}{c}{$r_e$} &  \multicolumn{1}{c}{$\langle\mu_e\rangle$} & \multicolumn{1}{c}{$A_{_{K_s}}$}  & \multicolumn{1}{c}{$z_{_{\rm{helio}}}$} & \multicolumn{1}{c}{$\log \sigma$} \\ 
        \multicolumn{1}{c}{(1)} & \multicolumn{1}{c}{(2)}  & \multicolumn{1}{c}{(3)}  & \multicolumn{1}{c}{(4)} & \multicolumn{1}{c}{(5)} & \multicolumn{1}{c}{(6)} & \multicolumn{1}{c}{(7)}           \\ \hline
WKK5920	&	9.93$\pm$0.04	&	4.63$\pm$0.43	&	15.03$\pm$0.21	&	0.086	&	0.0159&	2.312$\pm$0.012\\
WKK5972	&	9.64$\pm$0.05	&	6.72$\pm$0.76	&	15.54$\pm$0.25	&	0.079	&	0.0185&	2.414$\pm$0.010\\
WKK6012	&	10.89$\pm$0.05	&	4.48$\pm$0.38	&	15.94$\pm$0.19	&	0.080	&	0.0146&	2.172$\pm$0.014\\
WKK6019	&	9.94$\pm$0.04	&	3.72$\pm$0.26	&	14.56$\pm$0.15	&	0.070	&	0.0186&	2.409$\pm$0.010\\
WKK6047	&	11.90$\pm$0.07	&	3.07$\pm$0.21	&	16.11$\pm$0.16	&	0.070	&	0.0180&	2.010$\pm$0.015\\
WKK6116	&	9.39$\pm$0.03	&	6.66$\pm$0.56	&	15.33$\pm$0.19	&	0.066	&	0.0129&	2.344$\pm$0.012\\
WKK6180	&	10.15$\pm$0.05	&	6.07$\pm$0.59	&	15.87$\pm$0.22	&	0.065	&	0.0153&	2.308$\pm$0.011\\
WKK6183	&	10.30$\pm$0.04	&	4.35$\pm$0.30	&	15.27$\pm$0.16	&	0.063	&	0.0198&	2.377$\pm$0.012\\
WKK6198	&	11.71$\pm$0.09	&	4.29$\pm$0.58	&	16.67$\pm$0.31	&	0.056	&	0.0158&	1.905$\pm$0.020\\
WKK6204	&	9.39$\pm$0.04	&	6.58$\pm$0.64	&	15.28$\pm$0.21	&	0.065	&	0.0154&	2.499$\pm$0.009\\
WKK6221	&	10.70$\pm$0.06	&	6.44$\pm$0.48	&	16.51$\pm$0.17	&	0.066	&	0.0195&	2.038$\pm$0.017\\
WKK6229	&	11.59$\pm$0.04	&	1.92$\pm$0.07	&	14.78$\pm$0.09	&	0.063	&	0.0177&	2.210$\pm$0.015\\
WKK6233	&	11.77$\pm$0.06	&	2.36$\pm$0.16	&	15.42$\pm$0.16	&	0.062	&	0.0171&	2.243$\pm$0.014\\
WKK6235	&	10.95$\pm$0.05	&	3.31$\pm$0.30	&	15.36$\pm$0.21	&	0.069	&	0.0135&	2.141$\pm$0.018\\
WKK6242	&	10.92$\pm$0.03	&	2.09$\pm$0.09	&	14.31$\pm$0.10	&	0.061	&	0.0175&	2.424$\pm$0.011\\
WKK6250	&	10.47$\pm$0.04	&	3.18$\pm$0.24	&	14.74$\pm$0.17	&	0.064	&	0.0206&	2.324$\pm$0.010\\
WKK6269	&	8.27$\pm$0.05	&	15.95$\pm$1.44	&	16.07$\pm$0.20	&	0.060	&	0.0182&	2.579$\pm$0.011\\
WKK6282	&	11.23$\pm$0.04	&	1.56$\pm$0.17	&	14.00$\pm$0.24	&	0.060	&	0.0165&	2.274$\pm$0.012\\
WKK6305	&	9.15$\pm$0.04	&	8.05$\pm$0.61	&	15.45$\pm$0.17	&	0.080	&	0.0165&	2.327$\pm$0.009\\
WKK6318	&	8.70$\pm$0.06	&	15.52$\pm$1.46	&	16.48$\pm$0.21	&	0.073	&	0.0114&	2.354$\pm$0.013\\
WKK6342	&	11.04$\pm$0.03	&	2.39$\pm$0.06	&	14.72$\pm$0.06	&	0.062	&	0.0162&	2.326$\pm$0.009\\
WKK6360	&	10.11$\pm$0.02	&	3.14$\pm$0.08	&	14.35$\pm$0.06	&	0.063	&	0.0208&	2.505$\pm$0.009\\
WKK6383	&	10.95$\pm$0.05	&	4.00$\pm$0.27	&	15.74$\pm$0.15	&	0.071	&	0.0185&	2.195$\pm$0.012\\
WKK6431	&	10.87$\pm$0.03	&	2.50$\pm$0.16	&	14.68$\pm$0.14	&	0.075	&	0.0118&	2.286$\pm$0.010\\
WKK6473	&	11.94$\pm$0.07	&	2.00$\pm$0.16	&	15.23$\pm$0.19	&	0.067	&	0.0186&	2.064$\pm$0.017\\
WKK6477	&	11.84$\pm$0.06	&	2.69$\pm$0.21	&	15.79$\pm$0.18	&	0.073	&	0.0135&	2.106$\pm$0.019\\
WKK6555	&	11.04$\pm$0.06	&	2.79$\pm$0.38	&	15.05$\pm$0.30	&	0.067	&	0.0165&	2.207$\pm$0.010\\
WKK6600	&	10.07$\pm$0.05	&	5.15$\pm$0.58	&	15.41$\pm$0.25	&	0.068	&	0.0168&	2.338$\pm$0.011\\
WKK6615	&	12.04$\pm$0.08	&	2.22$\pm$0.21	&	15.59$\pm$0.22	&	0.060	&	0.0139&	2.108$\pm$0.021\\
WKK6620	&	12.94$\pm$0.21	&	2.83$\pm$0.56	&	16.97$\pm$0.48	&	0.062	&	0.0208&	1.777$\pm$0.054\\
WKK6679	&	10.79$\pm$0.04	&	3.19$\pm$0.20	&	15.10$\pm$0.14	&	0.074	&	0.0155&	2.159$\pm$0.012\\
\hline
\end{tabular}
\end{center}
\begin{flushleft}
\small{NOTES: 
%
The columns refer to (1) galaxy name  (2) the measured total extrapolated magnitude corrected for the star-subtraction and background effects (3) ef\mbox{}fective radius in arcsec corrected for the seeing effect (4) mean ef\mbox{}fective surface brightness in $\rm{mag}\;\rm{arcsec}^{-2}$, corrected for galactic extinction, redshift, and the cosmological dimming ef\mbox{}fects (5) galactic extinction from SF11: $A_{K_s} = 0.085\,A_B$ (6) galaxy redshift (heliocentric) (7) central velocity dispersion in dex, with both aperture correction and run offset applied.
}
\end{flushleft}

\end{table*}

\section{Relative distance between Norma and Coma} \label{offset}

\subsection{Fitting the Fundamental Plane} \label{fplane}

The FP relation (or the closely equivalent D$_n$-$\sigma$ relation)
has been widely used to determine the relative distances of ETGs
and measure peculiar velocities, e.g., \citep{Lynden_88, Lucey_88},
SMAC \citep{Hudson_99, Hudson_04}, ENEAR\footnote{Nearby Early-type Galaxies Survey} \citep{daCosta_00}, EFAR\footnote{Ellipticals FAR away} \citep{Colless_01}, 6dFGSv\footnote{Six-Degree Field Galaxy and Peculiar Velocity Survey} \citep{Magoulas_12}.

We have used the Coma cluster for calibration. To minimise the effect of sample selection biases and the effect of outliers, we fitted for the FP parameters by minimising the absolute residuals along the $\log \sigma$ direction \citep{Strauss_95, Jorgensen_96, LaBarbera_10}. Simultaneous least-squares fitting was used by constraining the FP parameters $a$ and $b$ to be the same for both Norma and Coma, while we allowed the FP intercepts to vary between the two clusters. 

\begin{figure}
 \centering
   \includegraphics[width=0.45\textwidth]{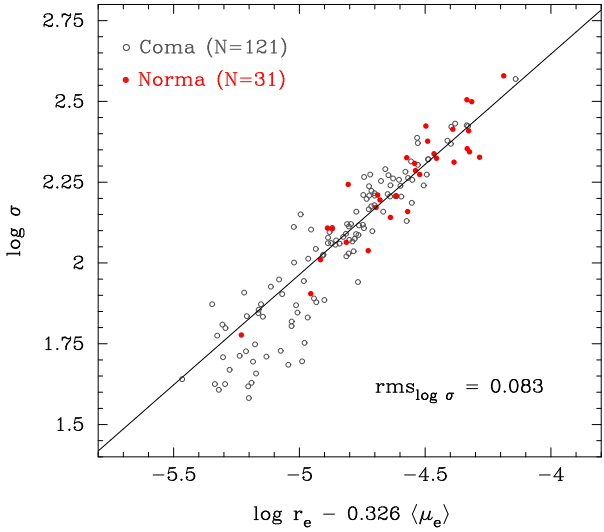} 
\caption{\label{FP_sigma} $K_s$-band FP projection, obtained by simultaneously minimising absolute residuals along $\log\sigma$; for the Coma cluster (open circles), and the Norma cluster (red filled circles). The Norma cluster has been shifted by applying the FP offset so that it lies at the Coma distance.}
\end{figure}

Our FP fit parameters are $a=1.465\pm0.059$, $b=0.326\pm0.020$, with an rms scatter of $\sim$\,0.08 dex in $\log \sigma$. The zero-point offset is $0.154$\,$\pm$\,$0.014$. 
Figure~\ref{FP_sigma} shows the projected FP. The Norma cluster which is represented by the red filled circles has been shifted to the Coma distance.
The error in individual galaxy distances is $\Delta_g = a\ln(10)\Delta_\sigma$. For cluster distances, the percentage error reduces according to the number of galaxies ($N$) in the sample, and is given by $\Delta_g / \sqrt{N}$. However, one needs to correct for the effect of Malmquist bias. The homogeneous Malmquist bias increases with distance but also decreases with the number of galaxies in the sample. For a measured distance $d$, the distance corrected for the homogeneous Malmquist bias is given by $d \exp(3.5\Delta_g^2/N)$ \citep{Hudson_97}. The correction is thus 0.59\% ($\sim$\,29~km\;s$^{-1}$ at the Norma distance). 

\subsection{The MIST algorithm}

For comparison, we have fitted the FP using the Measurement errors and Intrinsic Scatter Three dimensional (MIST) algorithm kindly provided by \citet*{LaBarbera_00}. The MIST algorithm is a bisector least-squares fit, used to determine the FP parameters $a$, $b$, and $c$. The statistical errors on the FP coefficients are computed through an inbuilt bootstrap analysis. For the MIST algorithm, we fitted the Coma FP (along $\log \sigma$-direction) and fixed the slopes $a$ and $b$ to determine the median value of the intercept, $c_{_{\rm N}}$ for the Norma cluster. The FP zero-point offset measured using this method is $0.154$\,$\pm$\,$0.019$. 

While low velocity dispersion galaxies have larger measurement errors (see Fig.~\ref{projFP}), we found no significant change in the derived Norma distance as a result of including these low velocity dispersion galaxies in our FP analysis. Through a bootstrap analysis (using the Coma ETGs), we analysed the change in the FP fit parameters with and without a magnitude cut. We applied a magnitude cut of 12\!\fmg5 as this effectively excludes most of the Coma ETGs with $\log \sigma < 2$ (see Section~\ref{fp_variables}). The bootstrap results (FP fit parameters) are shown in Fig.~\ref{bootstrap}. 

\subsection{Fixed metric aperture magnitudes: the modified Faber--Jackson relation}

Relative distances can also be measured using the Faber-Jackson (FJ) relation \citep{Faber_76} with galaxy magnitudes determined within a fixed metric radius \citep{Lucey_86}. This approach bypasses the uncertainties that arise from the determination of total magnitudes and allows relative distances to be measured to a similar accuracy to the FP. We have applied this technique as an alternative way to determine the Norma-Coma relative distance.

To make the \psf-corrections to the aperture photometry managable, we adopt a metric radius of 2\,kpc which corresponds, for our Coma cluster distance, to an aperture radius of 4\,\farcs16.
\galfit\ was used to determine the \psf-corrected aperture magnitudes for all galaxies in our Coma sample. For each galaxy in the Norma sample we determined a set of \psf-corrected aperture magnitudes that spanned a range of possible Norma cluster distances; if Norma has zero peculiar velocity then the 2\,kpc radius corresponds to a size of 5\,\farcs99. Galactic extinction and k-corrections were applied.

\begin{figure}
\centering
 \includegraphics[width=0.45\textwidth]{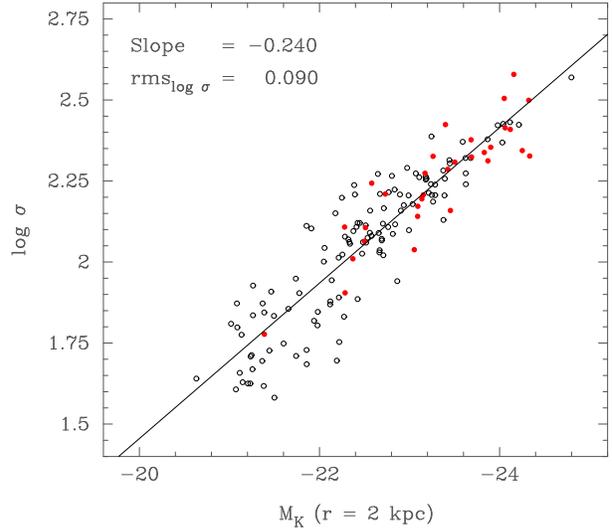} 
  \caption[Faber--Jackson relation]{\label{Faber_Jack}The FJ relation, based on aperture photometry using an aperture radius equal to 2\;kpc. The black open circles represent the Coma cluster, the red filled circles are from the Norma cluster.}
\end{figure}

A least-squares fit, minimising in the $\log \sigma$ direction, was used to simultaneously determine the slope of the combined L(r\,=\,2\,kpc)~--~$\sigma$ relation and the relative Norma-Coma offset. A range of Norma-Coma offsets were considered and for each the appropriate aperture size (i.e., based on the assumed relative Norma-Coma distance) was used for the Norma photometry. The best-fit was taken to be where there was no systematic difference in the  L(r\,=\,2\,kpc)~--~$\sigma$ relation between the two clusters and this corresponds to an offset of $0.159$\,$\pm$\,$0.022$ dex (see Fig.~\ref{Faber_Jack}). This derived value is in very good agreement with that directly determined from the FP analysis.

\section{The distance and peculiar velocity of the Norma cluster} \label{distances}

The measured Norma-Coma offsets derived from the three
different methods presented in Section~4 are in excellent agreement, i.e., 0.154 $\pm$ 0.014, 0.154 $\pm$ 0.019,  0.159 $\pm$ 0.022.
For our following analysis
we adopt the offset from the simultaneous least-squares fit
which has the smallest measurement error. This offset is
directly related to the difference in the angular diameter distances
(${D_A}$) of the two clusters, i.e.,
\begin{equation}
\log {D_A \left(\rm{Coma}\right)} - \log{D_A\left(\rm{Norma}\right)} =
0.154\,\pm\,0.014.
\label{finedist_here}
\end{equation}

In our analysis we assume that the Coma cluster has zero
peculiar velocity and hence Coma's
Hubble flow redshift, $z_{_{\rm{H}}}$, is equal to the
observed redshift in the local CMB rest frame, $z_{_{\rm{CMB}}}$.
For our adopted cosmology and a Coma CMB redshift
($z_{_{\rm{CMB}}}$) of 0.02400\,$\pm$\,0.00016
then $\log {D_A \left(\rm{Coma}\right)}$  = 1.996\,$\pm$\,0.003
where the angular diameter distance is given in Mpc.
Hence we find $\log {D_A\left(\rm{Norma}\right)}$ = 1.842\,$\pm$\,0.014
which implies a Hubble flow redshift for Norma ($z_{_{\rm{H}}}$)
of 0.01667 \,$\pm$\, 0.00055.

We derived Norma's peculiar velocity redshift, $z_{_{\rm PEC}}$,  via
$$ (\,1\,+\,z_{_{\rm CMB}})\,=\,(\,1\,+\,z_{_{\rm{H}}})(\,1\,+z_{_{\rm PEC}}) $$
\citep{Harrison_74}.
Norma's redshift in the local CMB rest frame,  $z_{_{\rm CMB}}$,
is 0.01652\,$\pm$\,0.00018.
Hence we derived a value for Norma's peculiar velocity of
--43\,$\pm$\,170\,km\,s$^{-1}$. Adding in the homogeneous Malmquist bias correction (see Section~\ref{fplane}) lowers this value to --72 \,km\,s$^{-1}$. Hence, we find that the Norma cluster has a small and insignificant peculiar velocity of --72\,$\pm$\,170\,km\,s$^{-1}$. 

\begin{figure}
 \centering
    \begin{tabular}{l c} 
         \includegraphics[width=0.44\textwidth]{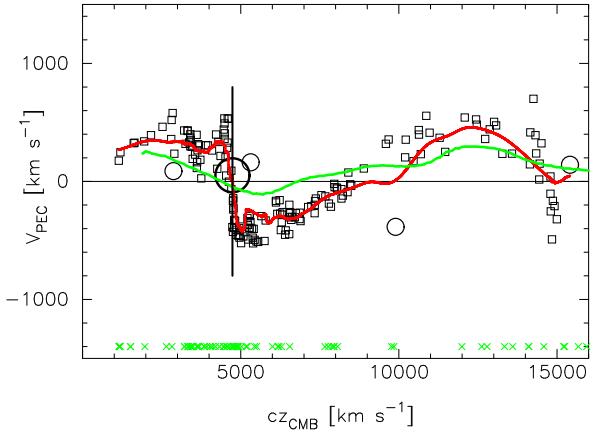}
\end{tabular}
\caption{\label{vp_models}Model predictions for the peculiar velocity in the Norma cluster direction. The green and red curves represent PSCz and 2MRS predictions, respectively. The crosses shown at the bottom of each panel show the redshifts of the PSCz galaxies in the cone. The black squares show the 2MRS predictions. The squares are non-grouped galaxies while the circles represent groups with the size of the circle reflecting the number of galaxies in the group. Norma is the large bold circle.
}
\end{figure}

All-sky galaxy redshift surveys map the local cosmography and
allow reconstructions of the density field (and the peculiar velocity field) to be made.
The reconstruction of the IRAS Point Source Catalogue redshift survey \citep[PSCz,][]{Saunders_00} by \citet{Branchini_99} provides a data cube of the predicted peculiar velocity
components $v_x$, $v_y$ and $v_z$ at grid positions $x$, $y$, $z$.
\citet{Lavaux_10}'s predictions derived from the 2MASS Redshift Survey \citep{Huchra_12}
are available at the Extragalactic Distance Database\footnote{\tt \url{edd.ifa.hawaii.edu/dfirst.php}}
\citep{Tully_09}. We have used these two reconstructions to estimate
the peculiar velocity in the Norma line-of-sight direction (see Fig.~\ref{vp_models}). The PSCz reconstruction under-estimates the peculiar velocities as compared to the 2MRS reconstruction possibly due to the former being less sensitive to galaxy clusters (Branchini, private communication). Both these studies predict a small peculiar velocity for Norma, i.e., less than $\pm$\,100\,km\,s$^{-1}$, which is in good agreement with our measurement of --72\,$\pm$\,170\,km\,s$^{-1}$.

In Table~\ref{m_errors}, we presented a summary of possible sources of measurement and systematic errors and the applied corrections. We found through simulations, that the effect of star-subtraction is very small, $\sim$\,$0\!\fmg 011$ which translates into a distance offset of $\sim$\,44\;km\;s$^{-1}$. The systematic effects arising from possible gradients in the sky background and non-resolved stars are very small, i.e., $\sim$\,32\;km\;s$^{-1}$. 
For extinction corrections, the correction factor ranges from 70\% to 90\% (e.g., \citealt{Bonifacio_00, Yasuda_07, Schroder_07,Schlafly_11}) which implies an uncertainty in distance in the range of --63\,km\,s$^{-1}$ to 16\,km\,s$^{-1}$. Note also that a $\pm$\,100\,km\,s$^{-1}$ peculiar velocity for Coma would change our value for Norma by $\pm$\,69\,km\,s$^{-1}$.

\section{Discussion and conclusions} \label{sec_discuss}

The GA is now widely identified as the 
Hydra-Cen-Norma supercluster \citep{Courtois_12,
Shaya_13, Tully_13}. Since the original discovery of the large positive peculiar 
velocities in the Hydra-Centaurus region \citep{Lynden_88},
new measurements, particularly for clusters, in the general GA region 
have been made. In Table~\ref{tab:vpec} we summarise the results from 
the FP-based SMAC and ENEARc surveys, the Tully-Fisher SFI++ 
survey and our Norma measurement. Although the typical measurement
errors are between 150 and 400\,km\,s$^{-1}$, most GA clusters
are observed to have positive peculiar velocities with an ``average'' value of \mbox{$\sim$+300\,km\,s$^{-1}$}. %

Our peculiar velocity measurement of \mbox{--72\,$\pm$\,170\,km\,s$^{-1}$}
for the Norma cluster is lower than most values found
for other GA clusters and hence may indicate that
Norma lies close to the GA's ``core''. The
uncertainties on the cluster peculiar velocities
listed in Table~\ref{tab:vpec} are sufficiently large that 
we can not currently determine whether or
not Norma partakes in the general GA outflow.
However there is strong independent
support for the GA outflow from the Type Ia
supernovae data (see \citealt{Lucey_05}) and many studies (e.g., \citealt{Hudson_99}) have attributed a major part of this large scale outflow to the Shapley Supercluster 
that lies $\sim$10\,000\,km\,s$^{-1}$ more distant in this general direction. %
\citet{Bolejko_08} estimate that Shapley can cause the GA
to have a net peculiar velocity of \mbox{$\sim$\,+80~km\;s$^{-1}$.}
%

While our knowledge of the local cosmography in the GA direction has improved considerably from galaxy redshift surveys that probe close to the galactic plane \citep{Radburn_06, Jones_09, Huchra_12}, there are still aspects that are incomplete and some important clusters/groups belonging to the GA may still remain hidden. An example is the discovery of CIZA J1324.7-5736 \citep*{Ebeling_02}, the second richest cluster in the GA region \citep{Nagayama_06}. Dedicated surveys are now underway to map larger swaths of the local universe veiled by the Milky Way, including ground-based near-IR surveys \citep[see, e.g.,][]{Kraan_11}, and in the mid-IR using WISE \citep{Bilicki_14, Jarrett_13} and in the radio by targeting neutral hydrogen, notably with HIPASS and HIZOA \citep{Henning_99, Henning_00, Kraan_00, Schroder_09}. Judicious application of the mid-infrared Tully-Fisher relation is now being considered to study the peculiar motions of gas-rich galaxies in massive structures that inhabit or cross the ZoA \citep{Sorce_12, Lagattuta_13, Said_13}. And our study of Norma has demonstrated that, despite the challenges of the large galactic extinction and severe stellar contamination, FP distances can be derived reliably for ETGs that lie relatively close to the galactic plane. The exploitation of such techniques will be essential to gain a more complete understanding of the GA and other important large scale structures that comprise the hidden cosmic web.
%
%
\begin{table} \scriptsize
\centering
\caption{\label{tab:vpec}Peculiar velocities of Clusters/Groups in the GA region.
 $N_{\phi}$ is the angle on the sky between the cluster and Norma.
SMAC measurements from \citet{Hudson_04}, SFI++ measurements from \citet{Springob_07},
ENEARc measurements from \citet{Bernardi_02}.
}
\begin{tabular}{l@{\hskip 0.13in}c@{\hskip 0.1in}r@{\hskip 0.1in}c@{\hskip 0.13in}c@{\hskip 0.13in}r@{\hskip 0.13in}r@{\hskip 0.13in}l}
\hline 
\hline
Cluster     & $\ell$   & $b$ & $N_{\phi}$ & cz$_{_{\rm CMB}}$ & N~ &  $v_{_{\rm PEC}}$~~~~~ & Source \\
            & $^{\circ}$ & $^{\circ}$ & $^{\circ}$ & km\,s$^{-1}$ &   & km\,s$^{-1}$~~~~& \\
\hline

A1060 (Hydra)   & 270 &   26 & 62 & 4055 & 26 &  +254 $\pm$ 223 & SMAC \\
                &     &      &    &      & 39 &  --47 $\pm$ 168 & ENEARc\\
                &     &      &    &      & 21 & --422 $\pm$ 169 & SFI++\\
AS636 (Antlia ) & 272 &   19 & 58 & 3129 & 17 &  +292 $\pm$ 102 & SFI++ \\
AS639           & 281 &   11 & 47 & 6526 &  6 & +1615 $\pm$ 453 & ENEARc \\
N3557 group     & 282 &   21 & 50 & 3318 &  7 &  +281 $\pm$ 160 & SFI++\\
A3526A (Cen30)  & 302 &   22 & 37 & 3300 & 27 &  +351 $\pm$ 136 & SMAC\\
                &     &      &    &      & 21 &  +500 $\pm$ 153 & ENEARc\\
                &     &      &    &      & 23 &  +260 $\pm$ 124 & SFI++\\
AS714           & 303 &   36 & 47 & 3576 &  7 &  +559 $\pm$ 245 & ENEARc\\
A3537           & 305 &   31 & 43 & 5370 &  4 &  +482 $\pm$ 560 & SMAC\\
CIZA J1324.7-5736 & 308 & 5 &  21   & 5899 &\\
E508 group      & 309 &   39 & 48 & 3310 &  9 &  +382 $\pm$ 122 & SFI++\\
A3574 (K27)     & 317 &   31 & 39 & 4881 &  8 &  +487 $\pm$ 351 & SMAC\\
                &     &      &    &      & 10 &  +479 $\pm$ 310 & ENEARc\\
                &     &      &    &      & 13 &  --199 $\pm$ 210 & SFI++\\
AS753           & 319 &   26 & 34 & 4431 & 14 &  +376 $\pm$ 282 & SMAC\\
                &     &      &    &      & 18 &  +812 $\pm$ 204 & ENEARc\\
A3581           & 323 &   33 & 40 & 6714 &  8 &  +131 $\pm$ 533 & SMAC\\
{\bf A3627 (Norma)}   & {\bf 325} & {\bf ~--7} & {\bf ~0}  & {\bf 4954} & {\bf 31} & {\bf --72 $\pm$ 170} & {\bf This study}\\
AS761           & 326 &   32 & 39 & 7076 & 11 &  +332 $\pm$ 483 & SMAC\\
AS805 (Pavo-II) & 332 & --23 & 17 & 4266 &  9 &  +293 $\pm$ 326 & SMAC\\
                &     &      &    &      & 12 &  --18 $\pm$ 268 & ENEARc\\
                &     &      &    &      &  8 &  +304 $\pm$ 147 & SFI++\\
Pavo I group    & 334 & --36 & 30 & 4055 & 16 &  +473 $\pm$ 191 & SFI++\\
\hline
\end{tabular}
\end{table}

\section*{Acknowledgments}

TM acknowledges financial support from the Square Kilometre Array South Africa (SKA~SA). PAW, SLB, THJ and MB acknowledge financial support from the South African National Research Foundation (NRF) and the University of Cape Town. %
JRL acknowledges support from STFC via ST/I001573/1. %
Funding for SDSS-III has been provided by the Alfred P. Sloan Foundation, the Participating Institutions, the National Science Foundation, and the U.S. Department of Energy Office of Science. The SDSS-III web site is \url{http://www.sdss3.org/}. SDSS-III is managed by the Astrophysical Research Consortium for the Participating Institutions of the SDSS-III Collaboration including the University of Arizona, the Brazilian Participation Group, Brookhaven National Laboratory, Carnegie Mellon University, University of Florida, the French Participation Group, the German Participation Group, Harvard University, the Instituto de Astrofisica de Canarias, the Michigan State/Notre Dame/JINA Participation Group, Johns Hopkins University, Lawrence Berkeley National Laboratory, Max Planck Institute for Astrophysics, Max Planck Institute for Extraterrestrial Physics, New Mexico State University, New York University, Ohio State University, Pennsylvania State University, University of Portsmouth, Princeton University, the Spanish Participation Group, University of Tokyo, University of Utah, Vanderbilt University, University of Virginia, University of Washington, and Yale University. This publication has made use of the NASA/IPAC Extragalactic Data base (NED), and also data products from the 2MASS, a joint project of the University of Massachusetts and the Infrared Processing and Analysis Center California Institute of Technology, funded by the National Aeronautics and Space Administration and the National Science Foundation.

\begin{appendix}
\section{Photometric analysis}

Table~\ref{Coma_phot} represents the ETGs in the Coma sample. No corrections have been applied to the total magnitudes presented in the table.
\begin{table*} 
\begin{center}
\caption[Coma cluster photometry results]{\label{Coma_phot} Coma cluster photometry results from f\mbox{}itting and extrapolating the galaxy surface brightness prof\mbox{}iles.}
\vskip-1.5em\begin{tabular}{c r r c c c  }  \\ \hline\hline
%
\multicolumn{1}{c}{Identification}  & \multicolumn{1}{c}{Tot. mag} &  \multicolumn{1}{c}{$r_e$} &  \multicolumn{1}{c}{$ \langle \mu_e\rangle$} & \multicolumn{1}{c}{$z_{_{\rm{helio}}}$} & \multicolumn{1}{c}{$\log \sigma$} \\ 
        \multicolumn{1}{c}{(1)} & \multicolumn{1}{c}{(2)}  & \multicolumn{1}{c}{(3)}  & \multicolumn{1}{c}{(4)} & \multicolumn{1}{c}{(5)} & \multicolumn{1}{c}{(6)}          \\ \hline
2MASXJ13023273+2717443	&	13.48$\pm$0.15	&	3.23$\pm$0.59	&	17.83$\pm$0.42	&	0.0254&	1.709$\pm$0.055\\
2MASXJ13020552+2717499	&	13.51$\pm$0.14	&	2.21$\pm$0.52	&	17.04$\pm$0.53	&	0.0243&	1.836$\pm$0.033\\
2MASXJ13000623+2718022	&	12.82$\pm$0.07	&	4.46$\pm$0.46	&	17.86$\pm$0.23	&	0.0263&	1.748$\pm$0.038\\
2MASXJ12593730+2720097	&	14.27$\pm$0.24	&	2.29$\pm$0.58	&	17.89$\pm$0.60	&	0.0233&	1.640$\pm$0.076\\
2MASXJ13010615+2723522	&	13.83$\pm$0.17	&	2.15$\pm$0.41	&	17.28$\pm$0.45	&	0.0271&	1.809$\pm$0.041\\
2MASXJ12564777+2725158	&	13.48$\pm$0.12	&	2.24$\pm$0.27	&	17.03$\pm$0.29	&	0.0259&	1.618$\pm$0.042\\
2MASXJ12583209+2727227	&	12.57$\pm$0.05	&	1.87$\pm$0.20	&	15.74$\pm$0.23	&	0.0234&	2.056$\pm$0.014\\
2MASXJ12573614+2729058	&	12.15$\pm$0.04	&	2.05$\pm$0.04	&	15.52$\pm$0.06	&	0.0241&	2.210$\pm$0.009\\
2MASXJ12570940+2727587	&	11.16$\pm$0.02	&	3.02$\pm$0.02	&	15.37$\pm$0.02	&	0.0249&	2.304$\pm$0.008\\
2MASXJ13002689+2730556	&	12.52$\pm$0.05	&	2.40$\pm$0.21	&	16.21$\pm$0.20	&	0.0260&	2.023$\pm$0.014\\
2MASXJ12580974+2732585	&	13.28$\pm$0.10	&	2.95$\pm$0.27	&	17.45$\pm$0.22	&	0.0233&	1.727$\pm$0.042\\
2MASXJ12573584+2729358	&	10.90$\pm$0.02	&	3.57$\pm$0.02	&	15.47$\pm$0.02	&	0.0244&	2.321$\pm$0.009\\
2MASXJ12572435+2729517	&	9.07$\pm$0.00	&	21.19$\pm$0.20	&	17.51$\pm$0.02	&	0.0245&	2.431$\pm$0.008\\
2MASXJ12570431+2731328	&	13.63$\pm$0.15	&	3.17$\pm$0.49	&	17.92$\pm$0.37	&	0.0277&	1.872$\pm$0.041\\
2MASXJ12563418+2732200	&	11.61$\pm$0.03	&	3.59$\pm$0.07	&	16.20$\pm$0.05	&	0.0237&	2.208$\pm$0.009\\
2MASXJ13014841+2736147	&	12.82$\pm$0.07	&	2.41$\pm$0.14	&	16.51$\pm$0.15	&	0.0275&	1.846$\pm$0.022\\
2MASXJ13011224+2736162	&	12.53$\pm$0.05	&	4.37$\pm$0.31	&	17.53$\pm$0.16	&	0.0252&	1.729$\pm$0.039\\
2MASXJ13001914+2733135	&	11.79$\pm$0.03	&	3.90$\pm$0.24	&	16.59$\pm$0.14	&	0.0196&	2.030$\pm$0.013\\
2MASXJ12585812+2735409	&	12.06$\pm$0.04	&	2.12$\pm$0.10	&	15.53$\pm$0.11	&	0.0200&	2.166$\pm$0.011\\
2MASXJ12573284+2736368	&	10.62$\pm$0.01	&	3.76$\pm$0.09	&	15.33$\pm$0.05	&	0.0201&	2.379$\pm$0.008\\
2MASXJ13020106+2739109	&	12.70$\pm$0.07	&	3.02$\pm$0.24	&	16.91$\pm$0.19	&	0.0236&	1.805$\pm$0.026\\
2MASXJ13015375+2737277	&	10.05$\pm$0.01	&	5.49$\pm$0.07	&	15.54$\pm$0.03	&	0.0262&	2.423$\pm$0.008\\
2MASXJ12591030+2737119	&	12.34$\pm$0.05	&	2.75$\pm$0.25	&	16.38$\pm$0.20	&	0.0191&	2.078$\pm$0.014\\
2MASXJ12571682+2737068	&	12.48$\pm$0.05	&	1.18$\pm$0.33	&	14.64$\pm$0.61	&	0.0242&	2.208$\pm$0.009\\
2MASXJ12594713+2742372	&	11.14$\pm$0.02	&	3.66$\pm$0.04	&	15.74$\pm$0.03	&	0.0280&	2.130$\pm$0.010\\
2MASXJ12584742+2740288	&	11.07$\pm$0.02	&	6.77$\pm$0.11	&	17.00$\pm$0.04	&	0.0280&	2.273$\pm$0.009\\
2MASXJ12583157+2740247	&	12.90$\pm$0.07	&	1.80$\pm$0.15	&	15.99$\pm$0.20	&	0.0231&	2.104$\pm$0.012\\
2MASXJ12563420+2741150	&	13.82$\pm$0.21	&	2.09$\pm$0.53	&	17.24$\pm$0.59	&	0.0229&	1.798$\pm$0.029\\
2MASXJ13000626+2746332	&	11.98$\pm$0.04	&	4.21$\pm$0.15	&	16.94$\pm$0.09	&	0.0206&	2.061$\pm$0.013\\
2MASXJ12592491+2744198	&	12.27$\pm$0.04	&	1.96$\pm$0.13	&	15.58$\pm$0.15	&	0.0201&	2.159$\pm$0.012\\
2MASXJ12591348+2746289	&	11.90$\pm$0.03	&	3.41$\pm$0.11	&	16.38$\pm$0.08	&	0.0228&	2.071$\pm$0.014\\
2MASXJ12590821+2747029	&	11.13$\pm$0.02	&	3.45$\pm$0.08	&	15.63$\pm$0.06	&	0.0234&	2.314$\pm$0.009\\
2MASXJ12590745+2746039	&	11.98$\pm$0.04	&	2.25$\pm$0.06	&	15.56$\pm$0.07	&	0.0212&	2.223$\pm$0.010\\
2MASXJ12585766+2747079	&	13.12$\pm$0.09	&	3.41$\pm$0.37	&	17.60$\pm$0.25	&	0.0231&	1.582$\pm$0.055\\
2MASXJ12585208+2747059	&	11.65$\pm$0.03	&	3.34$\pm$0.09	&	16.11$\pm$0.06	&	0.0189&	2.175$\pm$0.011\\
2MASXJ12581922+2745437	&	13.48$\pm$0.13	&	1.86$\pm$0.40	&	16.68$\pm$0.49	&	0.0182&	1.872$\pm$0.031\\
2MASXJ12574616+2745254	&	12.15$\pm$0.04	&	5.02$\pm$0.34	&	17.49$\pm$0.15	&	0.0204&	1.696$\pm$0.042\\
2MASXJ13011761+2748321	&	10.88$\pm$0.01	&	3.69$\pm$0.03	&	15.53$\pm$0.02	&	0.0244&	2.274$\pm$0.009\\
2MASXJ13003334+2749266	&	13.53$\pm$0.12	&	3.44$\pm$0.31	&	18.00$\pm$0.23	&	0.0273&	1.625$\pm$0.060\\
2MASXJ13000551+2748272	&	11.89$\pm$0.03	&	3.30$\pm$0.10	&	16.30$\pm$0.07	&	0.0219&	2.067$\pm$0.013\\
2MASXJ12595489+2747453	&	13.74$\pm$0.13	&	1.47$\pm$0.35	&	16.36$\pm$0.53	&	0.0275&	1.658$\pm$0.041\\
2MASXJ12593697+2749327	&	13.62$\pm$0.15	&	1.36$\pm$0.41	&	16.13$\pm$0.68	&	0.0209&	1.927$\pm$0.023\\
2MASXJ12592936+2751008	&	11.41$\pm$0.02	&	2.29$\pm$0.02	&	15.02$\pm$0.03	&	0.0226&	2.387$\pm$0.009\\
2MASXJ12580349+2748535	&	12.03$\pm$0.04	&	2.03$\pm$0.14	&	15.37$\pm$0.15	&	0.0238&	2.216$\pm$0.009\\
2MASXJ12574728+2749594	&	12.11$\pm$0.04	&	2.16$\pm$0.05	&	15.63$\pm$0.06	&	0.0202&	2.118$\pm$0.012\\
2MASXJ12571778+2748388	&	12.86$\pm$0.08	&	3.62$\pm$0.25	&	17.46$\pm$0.17	&	0.0238&	1.710$\pm$0.038\\
2MASXJ13025272+2751593	&	11.29$\pm$0.02	&	2.82$\pm$0.10	&	15.33$\pm$0.08	&	0.0274&	2.186$\pm$0.009\\
2MASXJ13015023+2753367	&	11.67$\pm$0.03	&	2.64$\pm$0.02	&	15.59$\pm$0.03	&	0.0252&	2.290$\pm$0.010\\
2MASXJ12594610+2751257	&	12.08$\pm$0.04	&	2.48$\pm$0.08	&	15.84$\pm$0.08	&	0.0270&	2.090$\pm$0.011\\
2MASXJ12593789+2754267	&	11.49$\pm$0.03	&	2.93$\pm$0.11	&	15.62$\pm$0.09	&	0.0267&	2.261$\pm$0.009\\
2MASXJ12592016+2753098	&	12.48$\pm$0.05	&	3.12$\pm$0.35	&	16.78$\pm$0.25	&	0.0216&	1.831$\pm$0.021\\
2MASXJ12590459+2754389	&	12.73$\pm$0.08	&	2.06$\pm$0.13	&	16.13$\pm$0.16	&	0.0214&	2.044$\pm$0.012\\
2MASXJ12590791+2751179	&	11.46$\pm$0.03	&	2.78$\pm$0.11	&	15.50$\pm$0.09	&	0.0219&	2.258$\pm$0.009\\
2MASXJ12575059+2752454	&	12.50$\pm$0.05	&	3.03$\pm$0.10	&	16.72$\pm$0.08	&	0.0232&	2.013$\pm$0.017\\
2MASXJ12572169+2752498	&	12.92$\pm$0.09	&	2.44$\pm$0.30	&	16.66$\pm$0.28	&	0.0248&	1.685$\pm$0.037\\
2MASXJ13004285+2757476	&	12.11$\pm$0.04	&	3.05$\pm$0.22	&	16.31$\pm$0.17	&	0.0281&	2.060$\pm$0.012\\
2MASXJ13004737+2755196	&	12.54$\pm$0.06	&	3.09$\pm$0.20	&	16.76$\pm$0.15	&	0.0286&	1.944$\pm$0.016\\
2MASXJ13003975+2755256	&	11.18$\pm$0.02	&	4.40$\pm$0.03	&	16.20$\pm$0.02	&	0.0250&	2.240$\pm$0.009\\
2MASXJ13002798+2757216	&	12.26$\pm$0.05	&	2.28$\pm$0.06	&	15.86$\pm$0.08	&	0.0234&	2.121$\pm$0.011\\
\hline
\end{tabular}
\end{center}
\begin{flushleft}
\small{NOTES: 
%
The columns refer to (1) 2MASS galaxy name  (2) the measured total extrapolated magnitude (3) ef\mbox{}fective radius in arcsec corrected for the seeing effect (4) mean ef\mbox{}fective surface brightness in $\rm{mag}\;\rm{arcsec}^{-2}$, corrected for galactic extinction, redshift, and the cosmological dimming ef\mbox{}fects (5) galaxy redshift (heliocentric) (6) central velocity dispersion (in dex) -- aperture corrections have been applied.
}
\end{flushleft}

\end{table*}
\begin{table*} 
\begin{center}
\contcaption{}
\vskip-1.5em\begin{tabular}{c r r c c c  }  \\ \hline\hline
%
\multicolumn{1}{c}{Identification}  & \multicolumn{1}{c}{Tot. mag} &  \multicolumn{1}{c}{$r_e$} &  \multicolumn{1}{c}{$ \langle \mu_e\rangle$} & \multicolumn{1}{c}{$z_{_{\rm{helio}}}$} & \multicolumn{1}{c}{$\log \sigma$} \\ 
        \multicolumn{1}{c}{(1)} & \multicolumn{1}{c}{(2)}  & \multicolumn{1}{c}{(3)}  & \multicolumn{1}{c}{(4)} & \multicolumn{1}{c}{(5)} & \multicolumn{1}{c}{(6)}           \\ \hline
2MASXJ12595670+2755483	&	12.49$\pm$0.05	&	3.76$\pm$0.33	&	17.16$\pm$0.20	&	0.0257&	2.001$\pm$0.016\\
2MASXJ12594438+2754447	&	11.25$\pm$0.02	&	3.74$\pm$0.09	&	15.93$\pm$0.06	&	0.0224&	2.207$\pm$0.009\\
2MASXJ12594234+2755287	&	12.53$\pm$0.05	&	1.20$\pm$0.15	&	14.74$\pm$0.27	&	0.0230&	2.238$\pm$0.011\\
2MASXJ12594423+2757307	&	12.68$\pm$0.07	&	3.10$\pm$0.25	&	16.95$\pm$0.19	&	0.0230&	1.819$\pm$0.025\\
2MASXJ12593570+2757338	&	8.98$\pm$0.00	&	20.02$\pm$0.12	&	17.30$\pm$0.01	&	0.0239&	2.426$\pm$0.009\\
2MASXJ12590414+2757329	&	12.98$\pm$0.11	&	2.10$\pm$0.32	&	16.40$\pm$0.35	&	0.0235&	2.112$\pm$0.014\\
2MASXJ12565310+2755458	&	12.26$\pm$0.04	&	1.25$\pm$0.01	&	14.59$\pm$0.05	&	0.0204&	2.272$\pm$0.009\\
2MASXJ12562984+2756240	&	11.70$\pm$0.03	&	3.47$\pm$0.15	&	16.23$\pm$0.10	&	0.0221&	2.266$\pm$0.009\\
2MASXJ13012713+2759566	&	12.60$\pm$0.05	&	1.15$\pm$0.16	&	14.70$\pm$0.31	&	0.0255&	2.198$\pm$0.010\\
2MASXJ13005445+2800271	&	10.46$\pm$0.01	&	6.97$\pm$0.18	&	16.54$\pm$0.06	&	0.0166&	2.371$\pm$0.009\\
2MASXJ13003877+2800516	&	11.97$\pm$0.03	&	4.58$\pm$0.16	&	17.07$\pm$0.08	&	0.0254&	2.026$\pm$0.014\\
2MASXJ13000809+2758372	&	8.51$\pm$0.00	&	17.27$\pm$0.06	&	16.52$\pm$0.01	&	0.0215&	2.570$\pm$0.008\\
2MASXJ13000643+2800142	&	12.02$\pm$0.04	&	3.22$\pm$0.23	&	16.37$\pm$0.16	&	0.0242&	2.081$\pm$0.013\\
2MASXJ12594681+2758252	&	11.53$\pm$0.02	&	4.60$\pm$0.11	&	16.60$\pm$0.05	&	0.0314&	2.116$\pm$0.012\\
2MASXJ12593827+2759137	&	13.11$\pm$0.10	&	3.70$\pm$0.48	&	17.77$\pm$0.30	&	0.0227&	1.909$\pm$0.022\\
2MASXJ12592657+2759548	&	13.06$\pm$0.09	&	2.27$\pm$0.13	&	16.66$\pm$0.16	&	0.0222&	1.904$\pm$0.018\\
2MASXJ12592136+2758248	&	13.48$\pm$0.14	&	2.53$\pm$0.53	&	17.34$\pm$0.47	&	0.0202&	1.713$\pm$0.050\\
2MASXJ12590603+2759479	&	11.30$\pm$0.02	&	4.83$\pm$0.24	&	16.52$\pm$0.11	&	0.0256&	2.179$\pm$0.010\\
2MASXJ12583023+2800527	&	11.14$\pm$0.02	&	3.71$\pm$0.03	&	15.80$\pm$0.03	&	0.0238&	2.282$\pm$0.009\\
2MASXJ13025659+2804133	&	12.22$\pm$0.05	&	4.45$\pm$0.10	&	17.26$\pm$0.07	&	0.0259&	1.753$\pm$0.030\\
2MASXJ13024442+2802434	&	10.98$\pm$0.02	&	6.38$\pm$0.26	&	16.84$\pm$0.09	&	0.0208&	2.254$\pm$0.010\\
2MASXJ13004867+2805266	&	10.89$\pm$0.01	&	3.60$\pm$0.02	&	15.48$\pm$0.02	&	0.0232&	2.321$\pm$0.008\\
2MASXJ13002215+2802495	&	11.54$\pm$0.03	&	4.56$\pm$0.17	&	16.62$\pm$0.08	&	0.0273&	2.087$\pm$0.011\\
2MASXJ13001702+2803502	&	12.35$\pm$0.04	&	2.40$\pm$0.23	&	16.09$\pm$0.21	&	0.0205&	2.070$\pm$0.013\\
2MASXJ13001475+2802282	&	11.80$\pm$0.03	&	2.11$\pm$0.03	&	15.27$\pm$0.04	&	0.0191&	2.206$\pm$0.010\\
2MASXJ13001286+2804322	&	12.35$\pm$0.05	&	1.85$\pm$0.10	&	15.48$\pm$0.12	&	0.0250&	2.075$\pm$0.011\\
2MASXJ13000803+2804422	&	11.57$\pm$0.02	&	2.13$\pm$0.07	&	15.02$\pm$0.08	&	0.0241&	2.263$\pm$0.008\\
2MASXJ12595601+2802052	&	11.04$\pm$0.02	&	4.85$\pm$0.15	&	16.25$\pm$0.07	&	0.0272&	2.206$\pm$0.010\\
2MASXJ12593141+2802478	&	11.86$\pm$0.04	&	2.88$\pm$0.13	&	15.98$\pm$0.10	&	0.0231&	2.108$\pm$0.011\\
2MASXJ12591389+2804349	&	12.11$\pm$0.05	&	5.36$\pm$0.26	&	17.55$\pm$0.12	&	0.0260&	2.150$\pm$0.014\\
2MASXJ12564585+2803058	&	13.51$\pm$0.12	&	2.83$\pm$0.48	&	17.59$\pm$0.39	&	0.0231&	1.669$\pm$0.048\\
2MASXJ12563890+2804518	&	13.17$\pm$0.11	&	4.55$\pm$0.61	&	18.25$\pm$0.31	&	0.0270&	1.625$\pm$0.062\\
2MASXJ13014700+2805417	&	11.11$\pm$0.02	&	3.47$\pm$0.03	&	15.66$\pm$0.02	&	0.0194&	2.257$\pm$0.008\\
2MASXJ13004459+2806026	&	12.46$\pm$0.05	&	2.92$\pm$0.20	&	16.61$\pm$0.16	&	0.0220&	1.891$\pm$0.017\\
2MASXJ13003552+2808466	&	11.96$\pm$0.03	&	4.56$\pm$0.16	&	17.10$\pm$0.08	&	0.0182&	1.885$\pm$0.020\\
2MASXJ12595511+2807422	&	12.25$\pm$0.04	&	2.98$\pm$0.23	&	16.42$\pm$0.17	&	0.0252&	2.096$\pm$0.014\\
2MASXJ12590392+2807249	&	10.35$\pm$0.01	&	4.90$\pm$0.12	&	15.59$\pm$0.05	&	0.0265&	2.422$\pm$0.008\\
2MASXJ12585341+2807339	&	12.52$\pm$0.05	&	2.10$\pm$0.09	&	15.95$\pm$0.11	&	0.0234&	2.062$\pm$0.013\\
2MASXJ12583636+2806497	&	11.25$\pm$0.02	&	3.53$\pm$0.07	&	15.80$\pm$0.04	&	0.0227&	2.239$\pm$0.009\\
2MASXJ12574670+2808264	&	13.41$\pm$0.14	&	3.76$\pm$0.83	&	18.11$\pm$0.50	&	0.0214&	1.607$\pm$0.066\\
2MASXJ13021025+2811309	&	13.05$\pm$0.11	&	3.01$\pm$0.43	&	17.29$\pm$0.33	&	0.0191&	1.834$\pm$0.025\\
2MASXJ13012280+2811456	&	12.12$\pm$0.04	&	3.35$\pm$0.22	&	16.55$\pm$0.15	&	0.0254&	2.109$\pm$0.013\\
2MASXJ13001795+2812082	&	10.12$\pm$0.01	&	6.45$\pm$0.13	&	15.95$\pm$0.04	&	0.0284&	2.369$\pm$0.008\\
2MASXJ12592021+2811528	&	12.88$\pm$0.05	&	2.94$\pm$0.18	&	16.98$\pm$0.14	&	0.0316&	1.949$\pm$0.022\\
2MASXJ12581382+2810576	&	12.13$\pm$0.05	&	5.57$\pm$0.30	&	17.68$\pm$0.12	&	0.0240&	1.869$\pm$0.019\\
2MASXJ12574866+2810494	&	11.69$\pm$0.03	&	2.65$\pm$0.22	&	15.62$\pm$0.18	&	0.0241&	2.159$\pm$0.011\\
2MASXJ12572841+2810348	&	11.69$\pm$0.03	&	4.54$\pm$0.14	&	16.76$\pm$0.07	&	0.0272&	2.021$\pm$0.015\\
2MASXJ12563516+2816318	&	12.03$\pm$0.04	&	3.04$\pm$0.20	&	16.26$\pm$0.15	&	0.0243&	2.090$\pm$0.013\\
2MASXJ12592611+2817148	&	13.58$\pm$0.12	&	1.86$\pm$0.35	&	16.72$\pm$0.42	&	0.0267&	1.695$\pm$0.039\\
2MASXJ12584394+2816578	&	13.78$\pm$0.17	&	1.49$\pm$0.36	&	16.45$\pm$0.55	&	0.0254&	1.629$\pm$0.046\\
2MASXJ12582949+2818047	&	13.29$\pm$0.12	&	2.44$\pm$0.50	&	17.06$\pm$0.46	&	0.0197&	1.844$\pm$0.027\\
2MASXJ13024079+2822163	&	11.45$\pm$0.03	&	5.14$\pm$0.24	&	16.81$\pm$0.11	&	0.0247&	1.941$\pm$0.020\\
2MASXJ13021434+2821099	&	12.62$\pm$0.08	&	5.40$\pm$0.56	&	18.06$\pm$0.24	&	0.0288&	1.855$\pm$0.032\\
2MASXJ13020865+2823139	&	11.01$\pm$0.02	&	5.61$\pm$0.17	&	16.56$\pm$0.07	&	0.0253&	2.213$\pm$0.020\\
2MASXJ13010904+2821352	&	12.51$\pm$0.05	&	2.65$\pm$0.39	&	16.44$\pm$0.33	&	0.0230&	1.879$\pm$0.018\\
2MASXJ13005207+2821581	&	10.80$\pm$0.01	&	4.27$\pm$0.10	&	15.75$\pm$0.05	&	0.0255&	2.240$\pm$0.010\\
2MASXJ13004423+2820146	&	11.78$\pm$0.03	&	2.82$\pm$0.20	&	15.82$\pm$0.16	&	0.0264&	2.098$\pm$0.010\\
2MASXJ13003074+2820466	&	10.78$\pm$0.01	&	6.03$\pm$0.15	&	16.52$\pm$0.06	&	0.0199&	2.205$\pm$0.009\\
2MASXJ13023199+2826223	&	13.44$\pm$0.16	&	3.70$\pm$0.76	&	18.12$\pm$0.48	&	0.0199&	1.776$\pm$0.064\\
2MASXJ12575392+2829594	&	12.46$\pm$0.06	&	1.69$\pm$0.21	&	15.42$\pm$0.28	&	0.0244&	2.121$\pm$0.012\\
2MASXJ12593568+2833047	&	12.25$\pm$0.04	&	2.26$\pm$0.11	&	15.82$\pm$0.11	&	0.0253&	2.113$\pm$0.012\\
2MASXJ12565652+2837238	&	11.97$\pm$0.03	&	2.95$\pm$0.16	&	16.14$\pm$0.13	&	0.0219&	2.036$\pm$0.013\\
\hline
\end{tabular}
\end{center}

\end{table*}
\subsection{Photometric analysis: star-subtraction} 
\begin{figure}
 \centering
 \begin{tabular}{l c} 
   \hspace{1.2cm}
   \includegraphics[width=0.125\textwidth]{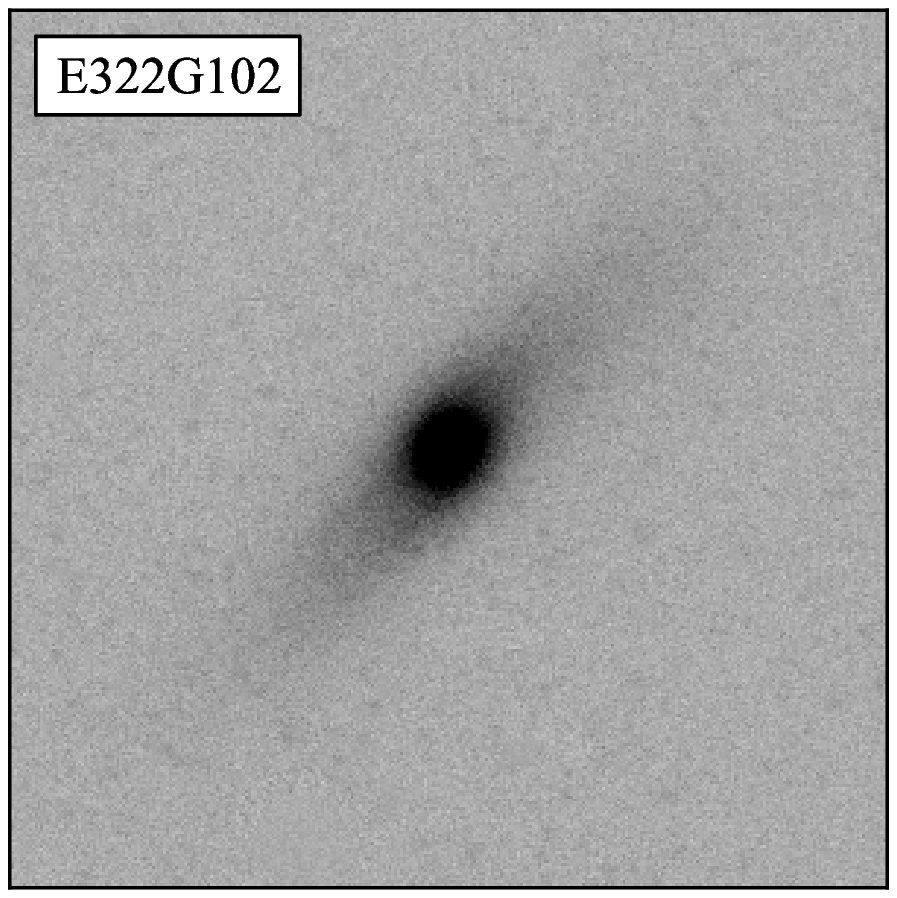}
\includegraphics[width=0.125\textwidth]{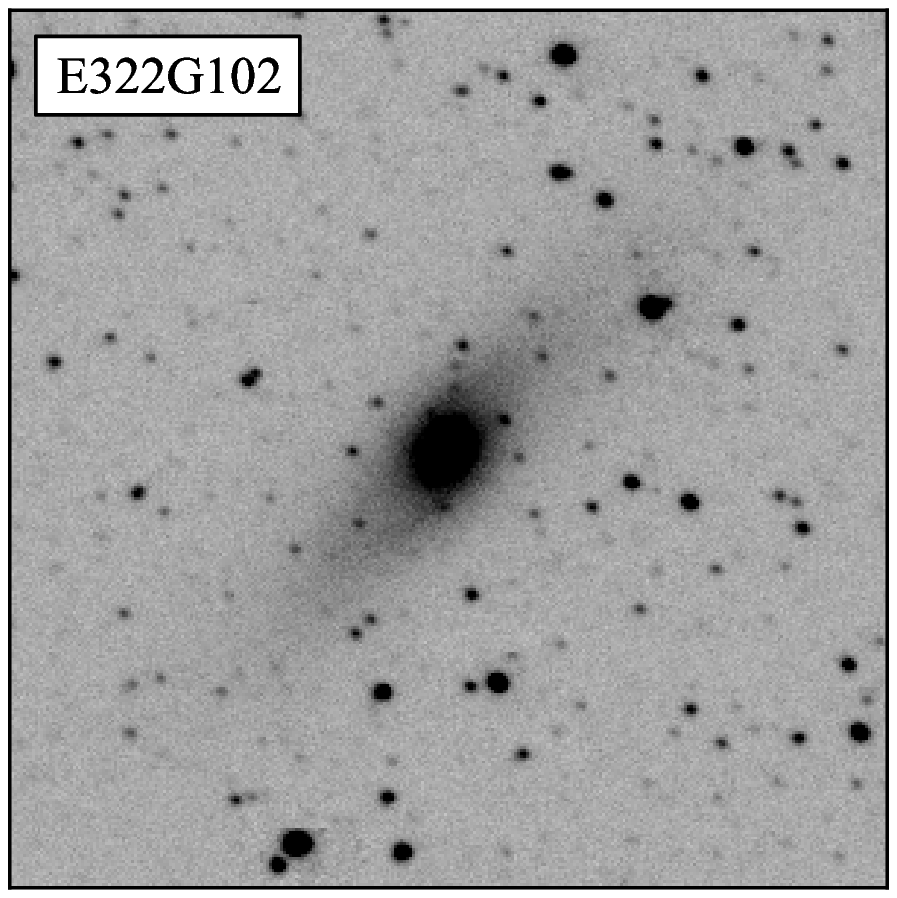}
\includegraphics[width=0.125\textwidth]{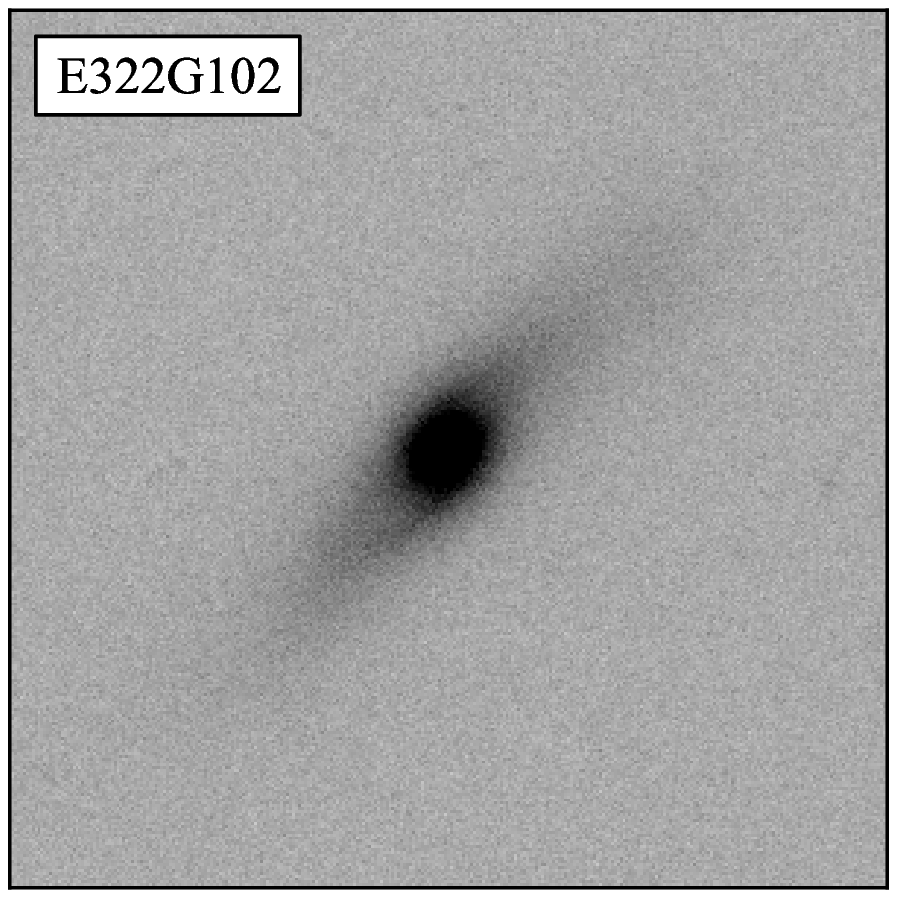} \\
     \includegraphics[width=0.46\textwidth]{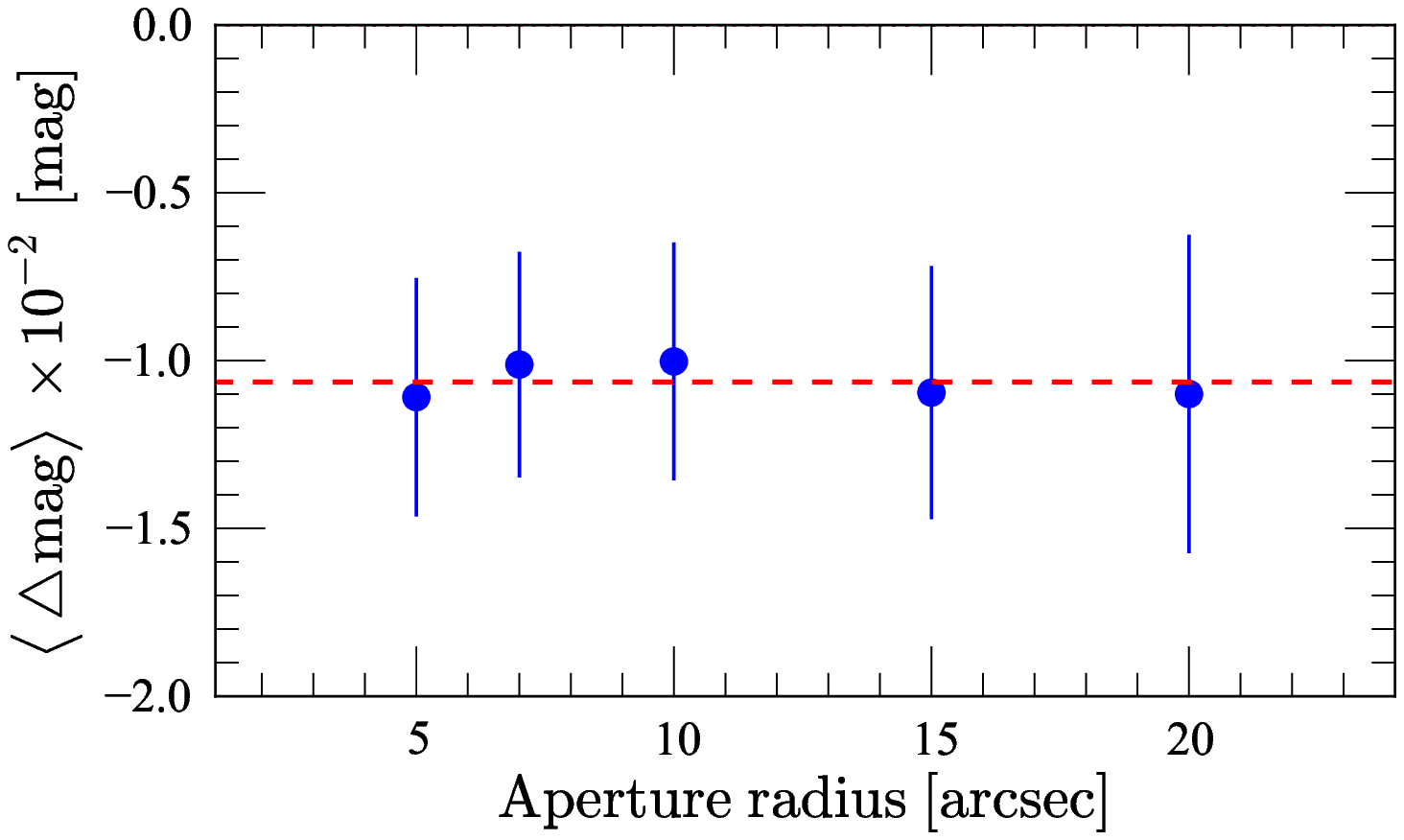} 
       \end{tabular}
\caption{
\label{av:mags}
The effect of star-subtraction, a comparison through aperture photometry. The average correction due to star-subtraction effect is $\bigtriangleup \text{m} = -0\!\fmg 0106 \pm 0\!\fmg 0003$.}
\end{figure}

Figure~\ref{av:mags} shows the simulation results, used to analyse the effect of star-subtraction on the photometric results of Norma ETGs. We used 12 Centaurus galaxies in the simulation. The top panels represent the original Centaurus image (left), the original image superimposed with stars from a typical Norma field (middle) and the star-subtracted image (right). The bottom panel represents the photometric results (aperture photometry). The y-axis is the average difference in the aperture magnitude per aperture radius over 12 Centaurus images, before and after adding and subtracting the added stars. 
The average difference as indicated by the red dashed line in Fig.~\ref{av:mags} is \mbox{$-0\!\fmg 0106 \pm 0\!\fmg0003$}. The scatter is only $\sigma=0.0006$. 

\section{Distribution of FP fit parameters} \label{fp_variables}

\subsection{FP projections} 
\begin{figure}
 \centering
 \includegraphics[width=0.48\textwidth]{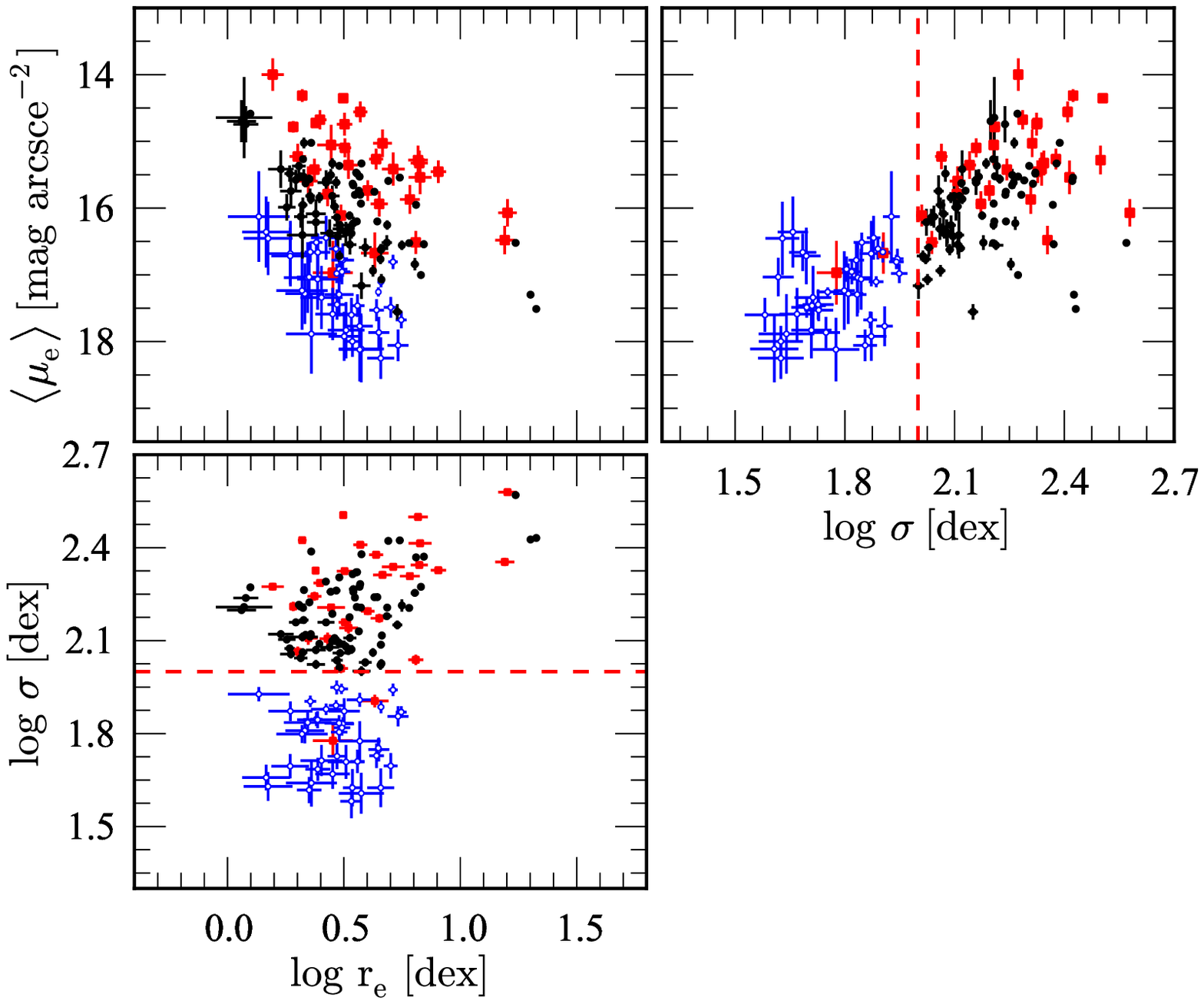}
\caption{\label{projFP} FP projections; the top left panel shows the Kormendy relation. The galaxies in the Norma cluster sample are represented by the red filled squares. The Coma ETGs have been subdivided into two: ETGs with central velocity dispersions less than $100\,\rm{km}\,s^{-1}$ are represented by the blue open circles, while the black filled circles represent Coma ETGs with $\log\sigma\ge2$.} 
\end{figure}

Figure~\ref{projFP} shows different FP projections showing the distribution of the ETGs in Norma (red filled squares) and Coma cluster (in blue open and black filled circles). The blue open circles represent Coma cluster ETGs with $\log \sigma \lt 2$ while the black filled circles represent Coma cluster ETGs with $\log \sigma \ge 2$. The vertical and horizontal dashed red lines represent $\log\sigma=2$, only two galaxies in the Norma sample have $\log\sigma\lt2$. On the other hand, there are 42 Coma ETGs with $\log \sigma<2$ and 79 ETGs with $\log \sigma \ge 2$.

\begin{figure*}
\centering
\begin{tabular}{l c} 
\includegraphics[width=0.42\textwidth]{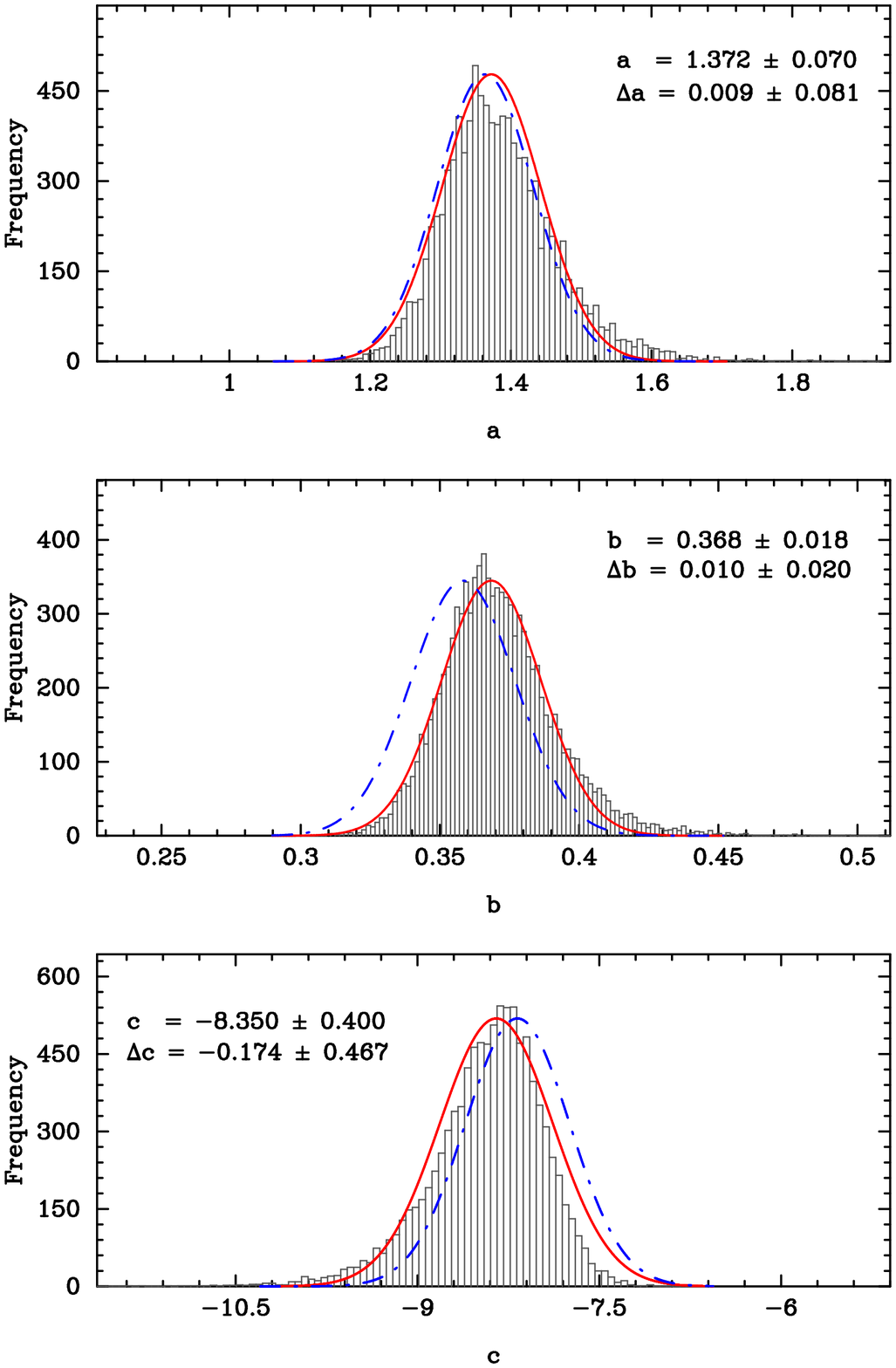} 
\hspace{0.5cm}
\includegraphics[width=0.42\textwidth]{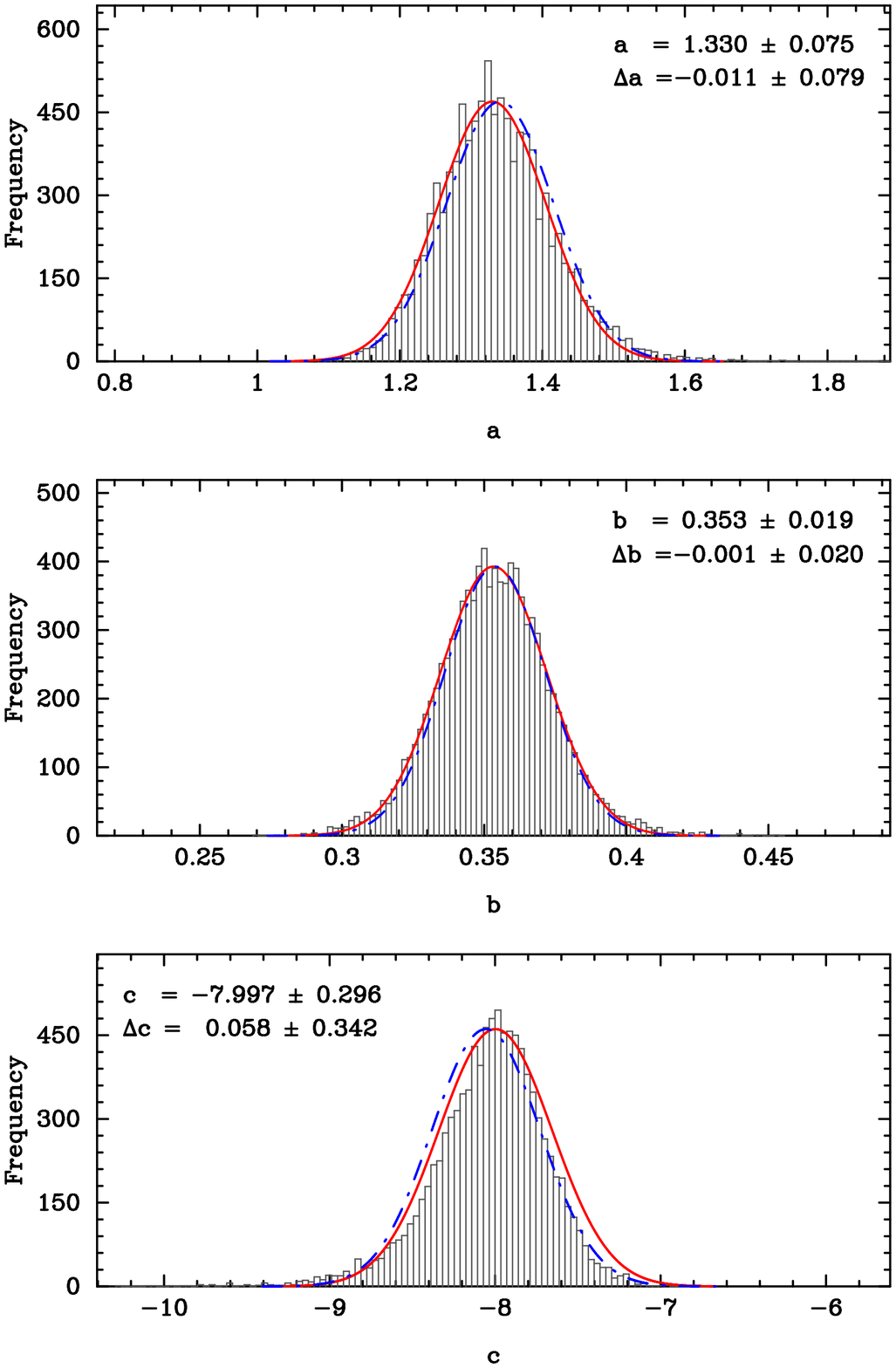}
\end{tabular}
\caption{\label{bootstrap} The distribution of the FP fit parameters through bootstrap re-sampling, using the MIST algorithm by minimising along the $\log\,\sigma$- direction. The Coma cluster galaxies were randomly selected, with replacement 10\,000 times. The left panels represent results when all the Coma ETGs were used while the right panels represent the bootstrap results for only Coma ETGs brighter than 12\fmg 5. The panels (top to bottom) show the distribution of the FP parameters ($a$, $b$, and $c$). The red curve is the Gaussian fit to the data. The blue curve represents the shift when the Gaussian fit parameters are replaced by FP fit parameters obtained by fitting the original Coma data -- the difference is indicated by $\Delta a$, $\Delta b$, and $\Delta c$.}
\label{lastpage}
\end{figure*}

Using the MIST algorithm, we checked the consistency of the FP fit parameters obtained by regressing along the $\log \sigma$ direction. By re-sampling the Coma cluster (with replacement) 10\,000 times, we analysed the distribution of the FP parameters $a$, $b$, and $c$. Figure~\ref{bootstrap} (left panels) shows the bootstrap results when all the 121 Coma ETGs were used while Fig.~\ref{bootstrap} (right panels) shows the results for only Coma ETGs brighter than 12\!\fmg5 ($N\!=\!78$, only 9\% of these have $\log\sigma\lt2$). The FP fit parameters in either case, are consistent with each other. The red solid curve is a Gaussian fit to the data. The mean value from the Gaussian fit of each of the FP parameters is indicated at the top part in each panel. The difference between the FP parameter from the original Coma data and that from the fitted distribution through bootstrap is also indicated (as $\Delta a$, $\Delta b$, and $\Delta c$).  The blue curve shows the small shift in the Gaussian fit due to this difference, i.e., $\Delta a$, $\Delta b$, and $\Delta c$. The magnitude cut of 12\!\fmg5 was motivated by the fact that it includes majority of Coma ETGs with $\log \sigma \gt 2$, thus making it possible to determine the effect (possible bias) of including ETGs with central velocity dispersions less than 100\;km\;s$^{-1}$, on the measured Norma distance. We found the zero-point offset to remain unchanged with and without the magnitude cut, i.e., the small shift $\Delta c$ in Fig.~\ref{bootstrap} (slightly larger in the left panels than the right panels) is accompanied by small changes in the Fundamental Plane fit parameters $\Delta a$, and $\Delta b$, which in turn affect the FP intercept of the Norma cluster, thereby leaving no significant changes in the zero-point offset.%

\end{appendix}

\end{document}